\documentclass{article}
\usepackage{etoolbox}
\let\bbordermatrix\bordermatrix
\patchcmd{\bbordermatrix}{8.75}{4.75}{}{}
\patchcmd{\bbordermatrix}{\left(}{\left[}{}{}
\patchcmd{\bbordermatrix}{\right)}{\right]}{}{}
\usepackage{authblk}  
\usepackage{lipsum}  
\usepackage[utf8]{inputenc}
\usepackage{natbib}
\bibpunct{(}{)}{;}{a}{}{;}
\usepackage{amsmath}  
\usepackage{amsfonts}  
\usepackage{graphicx}  
\usepackage{gensymb}  

\usepackage[table,xcdraw]{xcolor}
\usepackage{multirow}  
\usepackage{booktabs}  
\usepackage{makecell}
\usepackage{lscape}  
\usepackage[mathlines]{lineno}  
\usepackage{float}  
\usepackage{subfig}  
\usepackage{asymptote}
 \usepackage{fullpage} 
\usepackage{setspace}   
\usepackage{pdfpages}  

\usepackage{threeparttable}

\title{Characterising menotactic behaviours in movement data using hidden Markov models}
\author[1,2]{Ron R. Togunov}
\author[3]{Andrew E. Derocher}
\author[4]{Nicholas J. Lunn}
\author[1,5]{Marie Auger-Méthé}

\affil[1]{Institute for the Oceans and Fisheries, The University of British Columbia, Vancouver, BC V6T 1Z4, Canada}
\affil[2]{Department of Zoology, The University of British Columbia, Vancouver, BC V6T 1Z4, Canada}
\affil[3]{Department of Biological Sciences, University of Alberta, Edmonton, AB T6G 2E9, Canada}
\affil[4]{Wildlife Research Division, Science and Technology Branch, Environment and Climate Change Canada,CW-422 Department of Biological Sciences, University of Alberta, Edmonton, AB T6G 2E9, Canada}
\affil[5]{Department of Statistics, University of British Columbia, Vancouver, BC V6T 1Z4, Canada}

\date{\today\endgraf\bigskip 
\vspace{2cm}
\raggedright Corresponding Author: Ron R. Togunov \endgraf
Address: Institute for the Oceans and Fisheries, The University of British Columbia, Vancouver, BC V6T 1Z4, Canada\endgraf
Email: r.togunov@oceans.ubc.ca \endgraf
\vspace{1cm}
Running title: Menotactic hidden Markov model\endgraf
\vspace{1cm}
This is the pre-peer reviewed version of the following article: \textbf{Togunov, R. R., Derocher, A. E., Lunn, N. J., \& Auger-Méthé, M. (2021). Characterising menotactic behaviours in movement data using hidden Markov models. Methods in Ecology and Evolution, Accepted Article}, which has been published in final form at \textbf{https://doi.org/10.1111/2041-210X.13681}. This article may be used for non-commercial purposes in accordance with Wiley Terms and Conditions for Use of Self-Archived Versions.\endgraf
}

\begin{document}
\maketitle

\doublespacing
\newpage

\section{Abstract}
\begin{enumerate}
    \item Movement is the primary means by which animals obtain resources and avoid hazards. Most movement exhibits directional bias that is related to environmental features (\textit{taxis}), such as the location of food patches, predators, ocean currents, or wind. Numerous behaviours with directional bias can be characterized by maintaining orientation at an angle relative to the environmental stimuli (\textit{menotaxis}), including navigation relative to sunlight or magnetic fields and energy-conserving flight across wind. However, no statistical methods exist to flexibly classify and characterise such directional bias.
    \item We propose a biased correlated random walk model that can identify menotactic behaviours by predicting turning angle as a trade-off between directional persistence and directional bias relative to environmental stimuli without making \textit{a priori} assumptions about the angle of bias. We apply the model within the framework of a multi-state hidden Markov model (HMM) and describe methods to remedy information loss associated with coarse environmental data to improve the classification and parameterization of directional bias.
    \item Using simulation studies, we illustrate how our method more accurately classifies behavioural states compared to conventional correlated random walk HMMs that do not incorporate directional bias. We illustrate the application of these methods by identifying cross wind olfactory foraging and drifting behaviour mediated by wind-driven sea ice drift in polar bears (\textit{Ursus maritimus}) from movement data collected by satellite telemetry.
    \item The extensions we propose can be readily applied to movement data to identify and characterize behaviours with directional bias toward any angle, and open up new avenues to investigate more mechanistic relationships between animal movement and the environment. 
\end{enumerate}

Key words: behaviour, hidden Markov models, movement ecology, orientation, remote tracking, taxis, telemetry

\section{Introduction}
Behaviour is the primary way by which animals interact with their external environment to meet their needs. Nearly all biological activity manifests in the form of movement, from fine scale behaviours (e.g., grooming or prey handling) to large scale changes in location to access resources (e.g., food and mates) or avoid factors that increase energy expenditure or pose risk of injury \citep[e.g., predation or hazardous environments;][]{Wilmers2015TheEcology}. The improvement of tracking technology increases our ability to accurately identify and characterize behaviour, which is central to understanding animal ecology \citep{Kays2015TerrestrialPlanet,Wilmers2015TheEcology}. To answer key ecological questions, movement models should incorporate both internal conditions and environmental contexts to effectively leverage this wealth of data \citep{Kays2015TerrestrialPlanet, Schick2008UnderstandingDirections}. 

Animals tend to maintain heading (i.e., the orientation is autocorrelated), which can be modelled using a correlated random walk \citep[CRW;][]{Benhamou2006DetectingIndividual-dependent, Schick2008UnderstandingDirections}. When searching for resources in sparse environments, animals may alternate between different movement strategies. For example, in the absence of a desired resource, animals may enter an exploration phase (e.g., ranging or olfactory search), when they sense that they are near a target they may enter a localized exploitation phase (e.g., area restricted search), and when they reach the target, they may enter an exploitation phase \citep[e.g., grazing or prey handling;][]{Auger-Methe2016EvaluatingGuilds, Bartumeus2016ForagingUse, Schick2008UnderstandingDirections}. These types of movements are typically influenced or informed by the external environment. At the basic level, the environment may alter movement speed or trigger switching between different movement modes/behaviours without influencing directionality. For example, polar bears (\textit{Ursus maritimus}) may increase their time resting in low quality habitat to balance energy expenditure given prey availability \citep{Ware2017HabitatBears}. Movement orientation, measured as turning angle between successive locations, can also be affected by or modulated in response to environmental stimuli or cues \citep[e.g., foraging patches, nest sites, predators, and directions of ocean current, wind, and sunlight;][]{Benhamou2006DetectingIndividual-dependent,  Codling2008RandomBiology}. In the animal movement literature, preference for moving in a particular direction is defined as \emph{bias}. Directional response to the external stimuli is a special case of bias defined as \emph{taxis} \citep{Codling2008RandomBiology}. The degree of bias among different behaviours occurs along a spectrum from being primarily biased toward a preferred direction or an external factor (biased random walk; BRW), to a trade-off between directional persistence and directional bias (biased correlated random walk; BCRW), to being primarily governed by directional persistence \citep[CRW;][]{Benhamou2006DetectingIndividual-dependent, Codling2016BalancingMovement, Codling2008RandomBiology}. Research has explored movement where bias is directly toward (i.e., positive taxis) or away (i.e., negative taxis) from a target/focal point - for example, bias relative to light \citep[\textit{phototaxis}; e.g.,][]{Park2017PhototacticLEDs}, sound \citep[\textit{phonotaxis}, e.g.,][]{Diego-Rasilla2004HeterospecificMarmoratus}, and water currents \citep[\textit{rheotaxis}; e.g.,][]{Mauritzen2003FemaleTreadmill, Savoca2017OdoursFish}. 

\begin{figure}
    \centering
    \includegraphics{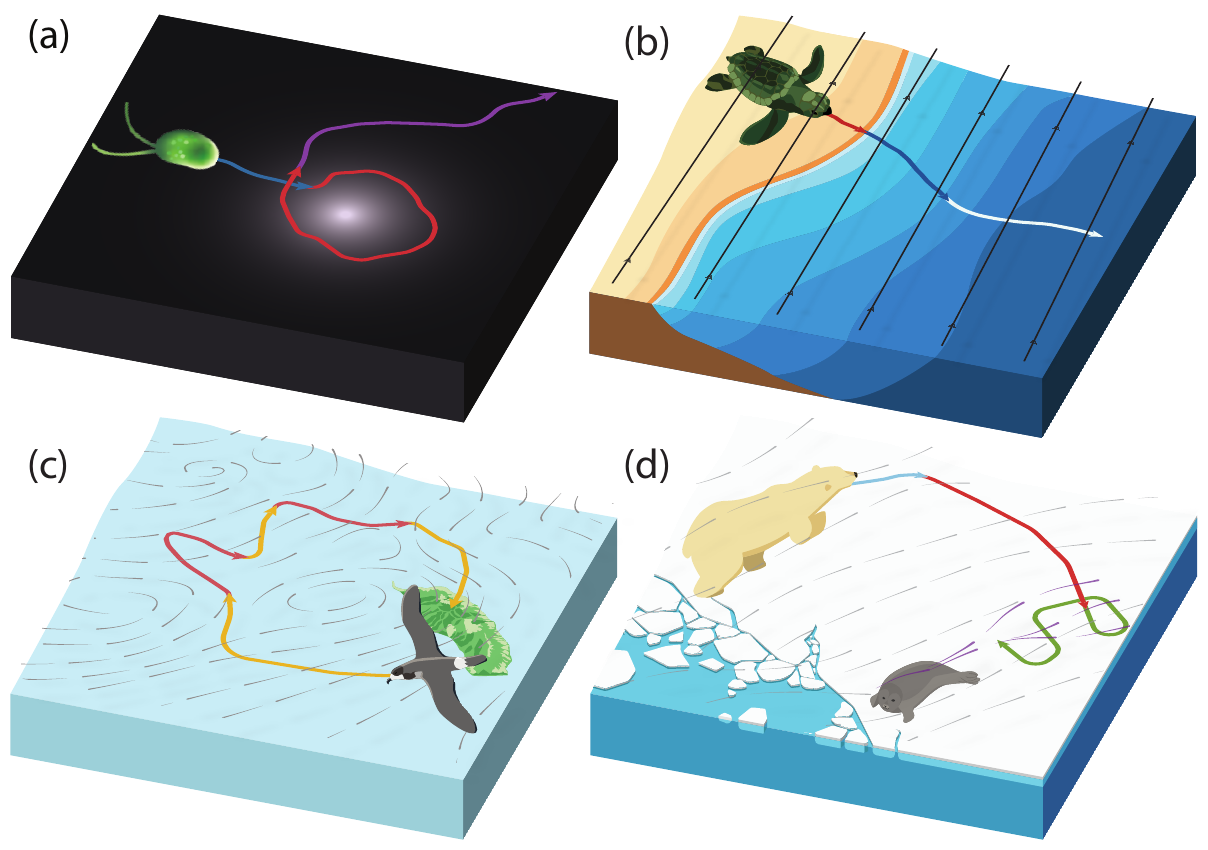}
    \caption{Examples menotaxis in animal movement: (a) Microalga positive \textit{phototaxis} toward light (blue), perpendicular to maintain constant light intensity (red), then away from light \citep[purple;][]{Arrieta2017}; (b) hatchling loggerhead sea turtles following visual cues toward brighter ocean (red), perpendicular orientation relative to waves to move away from shore (blue), and long-distance navigation guided by magnetic inclination (white; \textit{magnetotaxis}) \citep{Lohmann2008TheNavigation}; (c) Desertas petrel transitory flights (yellow) approximately $50\degree$ relative to wind (grey field) and $80\degree$ during olfactory search (red) \citep{Ventura2020GadflyScales}; and (d) polar bear track at $15\degree$ relative to wind when stationary on sea ice (blue), approximately $90\degree$ relative to wind during olfactory search (red) to maximize probability of encountering odour plumes (purple), and movement apparently random relative to wind during area restricted search (green) \citep[][]{Togunov2017WindscapesCarnivore, Togunov2018Corrigendum:10.1038/srep46332}.}
    \label{fig: biotaxis}
\end{figure}

In contrast to simple positive or negative taxis, some behaviours exhibit bias toward a constant angle relative to stimuli or target (\textit{menotaxis}). For example, the microalga \textit{Chlamydomonas reinhardtii} exhibits movement at an angle to light to maintain preferred luminosity \citep[Fig. \ref{fig: biotaxis}a;][]{Arrieta2017}. Loggerhead  sea  turtle (\textit{Caretta caretta}) hatchlings exhibit positive phototaxis to get to shore following nest emergence, followed by movement perpendicular to waves to move away from shore, then using magnetic fields for large-scale navigation \citep[\textit{magnetotaxis}; Fig. \ref{fig: biotaxis}b;][]{Lohmann2008TheNavigation, Mouritsen2018Long-distanceAnimals}. Some seabirds exhibit bias relative to wind direction (\textit{anemotaxis}); fly approximately 50\degree relative to wind to maximize ground speed during transitory flights, and fly crosswind to maximize the chance of crossing an odour plume during olfactory search \citep[Fig. \ref{fig: biotaxis}c;][]{Nevitt2008EvidenceExulans., Ventura2020GadflyScales}. Similar crosswind olfactory search has been observed in several other taxa \citep[Fig. \ref{fig: biotaxis}c,d;][]{Baker2018AlgorithmsSpecies, Kennedy1974Pheromone-regulatedMoths, Togunov2017WindscapesCarnivore,Togunov2018Corrigendum:10.1038/srep46332}. In mobile environments (i.e., aerial or aquatic systems), the observed motion determined from remote tracking reflects both voluntary movement and advection by the system, and also influences the apparent orientation of movement \citep{Auger-Methe2016HomeIce, Gaspar2006MarineTrack, Schick2008UnderstandingDirections}. For example, the motion of a stationary polar bear on sea ice reflects sea ice drift \citep{Auger-Methe2016HomeIce}, which tends to move $20\degree$ relative to wind, the primary driver of drift \citep[Fig. \ref{fig: biotaxis}d;][]{Bai2015ResponsesStudy}. 

Identifying behaviours in movement data is a field of active development, and there is a need to develop more sophisticated movement models that consider the perceptual and cognitive capacities of animals \citep{Auger-Methe2016EvaluatingGuilds, Bracis2017MemoryMigration, Gaynor2019LandscapesResponse, Kays2015TerrestrialPlanet}. Hidden Markov models (HMMs) are well-developed statistical models used to describe behavioural changes using movement data. HMMs and other movement models that integrate relationships between environmental data and movement characteristics (e.g., speed and orientation) have proven to be effective at elucidating interactions between movement and the environment \citep{Kays2015TerrestrialPlanet, Mcclintock2020UncoveringModels}. Thus far, HMMs have only been used to model positive and negative taxis on turning angle. The detection of biased behaviours may also be confounded by mismatched spatial and temporal resolutions among data streams. Mismatched, multi-stream, multi-scale data are widespread in the field of spatial ecology and there is a need for statistical tools to integrate disparate data sources \citep{Adam2019JointModels, Bestley2013IntegrativePredator, Fagan2013SpatialMovement, Wilmers2015TheEcology}. In this paper, we first present an extension to BCRW HMMs that relaxes the direction of bias and allows modelling movement toward any angle relative to stimuli (i.e., menotaxis). Second, we propose incorporating a one-step transitionary state in HMMs when faced with low resolution environmental data to improve characterizing the direction of bias in BRWs. We investigate the accuracy of our model in two simulation studies and illustrate its application using polar bear telemetry data as a case study. Finally, we provide a detailed tutorial for these methods with reproducible R code in Appendix \ref{appendix:D}.  

\section{Methods}

\subsection{Model formulation}
\subsubsection{Introduction to modelling CRW, BRW, and basic BCRW}

\begin{figure}  
    \centering
    \includegraphics[scale = 0.7]{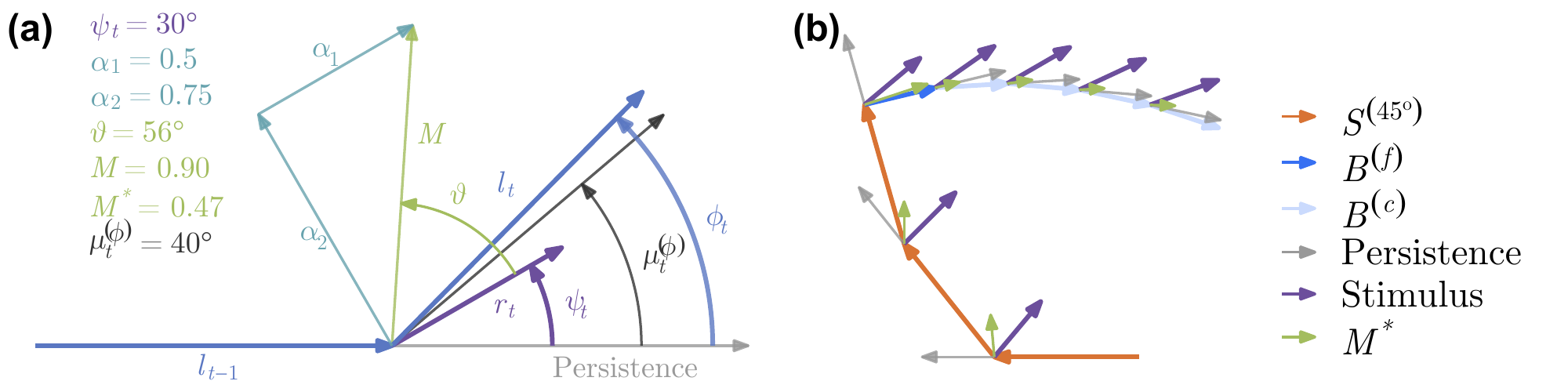}
    \caption{Illustration of the proposed menotactic BCRW for one step (a) and a sample three-state track (b). Panel (a) presents the notation used in this paper along with their values. Bold vectors represent values obtained from the data (animal track, blue vectors, and external stimulus, purple vector) and thin vectors represent values estimated by the model. $l_t$ and $l_{t-1}$ represent the step length, the light grey unit vector represents directional persistence from which turning angle $\phi_t$ is calculated. $\psi_t$ (purple arc) represents the angle of stimulus with magnitude $r_t$ relative to the bearing at $t-1$ (vector of persistence; light grey). The green arrow represents the vector of attraction with a magnitude of attraction $M$ and angle $\vartheta$ relative to stimulus, which are obtained from the estimated bias parallel, $\alpha_1$, and perpendicular, $\alpha_2$, to the stimulus (teal). $\mu_t^{(\phi)}$ (dark grey arc) represents the expected mean turning angle as a trade-off between persistence and bias toward $\vartheta$. $M^\ast$ represents the scaled magnitude of attraction. Panel (b) presents a sample track with a BCRW state with bias $\vartheta_{S^{(45\degree)}} = 45\degree$ left of stimulus ($S^{(45\degree)}$; orange). A second BRW behaviour ($B$) with bias toward $\vartheta_{B} = -22\degree$ relative to stimulus is divide into a BRW state for the first step ($B^{(f)}$; blue) and a CRW state for consecutive steps ($B^{(c)}$; light blue). Green vector represents the scaled magnitude of attraction $M^\ast$ with direction $\vartheta$ relative to the stimulus. To represent environmental error, the true direction of the stimulus (not shown) rotates $-10\degree$ each step, while the estimated stimulus (purple) rotates $-5\degree$ each step.}
     \label{fig:vectors_sample}
\end{figure}

Animal movement observed using location data (i.e., latitude and longitude) is typically described using two  data streams: step length $l_t \in (0, \infty)$ (distance between consecutive locations) and turning angle $\phi_t \in (-\pi,\pi]$ (change in bearing between consecutive steps), where $t \in 1, ..., T$ represents the time of a step (Fig. \ref{fig:vectors_sample}; the notation used in this paper is described in Table \ref{tab:notation}). The probability of the step length $l_t$ from location at time $t$ to $t+1$ is often modelled with a Weibull or gamma distribution \citep{Langrock2012FlexibleExtensions,Mcclintock2018MomentuHMM:Movement}. We  assume that $l_t$ is distributed as follows:
\begin{linenomath*} 
\begin{equation} \label{eq:gamma_pdf}
l_t \sim \mathrm{gamma}(\mu_{t}^{(l)}, \sigma_{t}^{(l)}),\\
\end{equation}  
\end{linenomath*}
where $\mu_{t}^{(l)} \in (0, \infty)$ and $\sigma_{t}^{(l)} \in (0, \infty)$ are the mean and standard deviation of the step length (these parameters can also be derived from shape and scale parametrisation of the gamma distribution). Animal movement is often influenced by external factors that result in complex movement patterns \citep{Auger-Methe2016HomeIce}. For example, some animals reduce speed when travelling through deep snow, increase speed when flying downwind, modulate orientation towards a foraging patch, or moving away from predators \citep{Duchesne2015EquivalenceMovement, Mcclintock2018MomentuHMM:Movement, Mcclintock2012AWalks}. The influence of these factors on movement can be represented by modelling the movement parameters as functions of external factors \citep[e.g.,][]{Mcclintock2018MomentuHMM:Movement}. We can model behaviours where the mean speed of observed movement is associated with external factors using:

\begin{linenomath*}  \begin{align}  \label{eq:sl_m}
\mathrm{ln}(\mu_{t}^{(l)}) &=
\beta_{1} + \beta_{2} r_t,
\end{align}  \end{linenomath*}
where $\beta_{1} \in (-\infty, \infty)$ is an intercept coefficient for step length mean and $\beta_{2} \in (-\infty, \infty)$ is a slope coefficient representing how the mean step length is affected by magnitude of external stimulus $r_t$ (e.g., wind speed, or speed of neighbouring individuals in a school of fish). 

\begin{table}[]
    \caption{Description of acronyms and notation used in this paper and their interval, if applicable.}
    \label{tab:notation}
    \centering
    \begin{tabular}{ccl}
        \toprule 
        Acronym/variable & Interval & Description  \\ \hline
        CRW &-& Correlated random walk.  \\
        BRW &-& Biased random walk. \\
        BCRW &-& Biased correlated random walk.  \\
        TBCRW &-& \makecell[l]{Biased correlated random walk incorporating \\ \ \ transitionary states.} \\
        $\pi$ & $\pi$ & The value 3.14159... \\
        $T$ & $\{1,2,...\}$ & Total number of time steps. \\
        $t$ & $[1, T]$ & A time step.  \\
        $X_t$ & - & Set of observations at time $t$. \\
        $\textbf{X} $ & - & The set of all observations $(X_1,..., X_T)$. \\
        $l$ & $[0,\infty)$ & Step length between consecutive locations. \\
        $\phi$ & $(-\pi,\pi]$ & \makecell[l]{Turning angle (i.e., change in bearing) between \\ \ \ consecutive steps.} \\
        $r$ & $[0, \infty)$ & Magnitude of the stimulus.  \\
        $\psi$ & $(-\pi,\pi]$ & \makecell[l]{Direction of a stimulus relative to the bearing of the \\ \ \ previous time step.} \\
        $\mu^{(l)}$ & $(0,\infty)$ & Mean parameter of step length. \\
        $\sigma^{(l)}$ & $(0,\infty)$ & Standard deviation parameter of step length. \\
        $\beta_1$ & $(-\infty, \infty)$ & Intercept coefficient for mean step length. \\
        $\beta_2$ & $(-\infty, \infty)$ & Slope coefficient for mean step length and $r$. \\
        $\mu^{(\phi)}$ & $(-\pi,\pi]$ & Mean parameter of turning angle.  \\
        $\kappa^{(\phi)}$ & $(0,\infty)$ & Concentration parameter of turning angle.  \\
        $\alpha_1$ & $(-\infty, \infty)$ & Bias coefficient in the same direction as the stimulus. \\
        $\alpha_2$ & $(-\infty, \infty)$ & Bias coefficient $90\degree$ left of the stimulus. \\
        $\vartheta$ & $(-\pi,\pi]$ & The direction of bias relative to stimulus. \\
        $M$ & $[0,\infty)$ & Magnitude of attraction. \\
        $M^\ast$ & $[0,1)$ & Scaled magnitude of attraction.  \\
        HMM &-& Hidden Markov model. \\
        $N$ & $\{1,2,...\}$ & Total number of behavioural states.  \\
        $S$ & $[1,N]$ & Behavioural state.  \\
        $\gamma_{i,j}$ & $[0,1]$ & Transition probability from state \textit{i} to state \textit{j}.  \\
        $\mathbf{\Gamma}$ &-& $N \times N$ Transition probability matrix.  \\
        $B, B^{(f)}, B^{(c)}$ &-& \makecell[l]{A BRW state, a state for the first step in the BRW, \\ \ \ and a state for consecutive BRW steps, respectively.}  \\
        $D, D^{(f)}, D^{(c)}$ &-& \makecell[l]{The drift state, a state for the first step in drift, \\ \ \ and a state for consecutive drift steps, respectively.}  \\
        $O^{(L)}, O^{(R)}$ &-& \makecell[l]{Olfactory search state with anemotaxis left of wind \\ \ \ and anemotaxis right of wind, respectively.}  \\
        $ARS$ &-& Area restricted search state.  \\
        \bottomrule
 \end{tabular}
\end{table}

The probability of turning angle $\phi_t$ is often modelled with a wrapped Cauchy or von Mises distribution \citep{Mcclintock2020UncoveringModels}. We assume that $\phi_t$ follows a von Mises distribution as follows:
\begin{linenomath*} \begin{equation}  
\phi_t \sim \mathrm{vMises}(\mu_{t}^{(\phi)}, \kappa_{t}^{(\phi)})
\label{eq:vonMisesPDF}
\end{equation}  \end{linenomath*}
where $\mu^{(\phi)}_{t} \in (-\pi,\pi]$ is the mean turning angle parameter at time $t$ and $\kappa^{(\phi)} \in (0, \infty)$ is the concentration parameter around $\mu_{t}^{(\phi)}$ (Fig. \ref{fig:vectors_sample}a). In a basic CRW, $\mu_{t}^{(\phi)}$ is assumed to equal zero and only the concentration parameter $\kappa^{(\phi)}$ is modelled. In BRW and BCRW with simple positive or negative taxis (i.e., bias toward or away from a target), the mean turning angle $\mu_t^{(\phi_t)}$ can be modelled as a function of the orientation relative to a stimulus $\psi_t$ \citep[e.g., direction of den site; Fig. \ref{fig:vectors_sample};][]{Mcclintock2018MomentuHMM:Movement}; we assume $\phi_t$ follows a circular-circular von Mises regression model based on \citet{Rivest2016AAnalysis}:
\begin{linenomath*} \begin{equation}  
    \label{eq:BCRW.TAm}
        \mu_{t}^{(\phi)} = \mathrm{atan2}(\alpha_{1} \sin \psi_t, 1 + \alpha_{1} \cos \psi_t) \\
\end{equation}  \end{linenomath*}
where $\psi_t \in (-\pi,\pi]$ is the observed angle of stimulus at $t$ relative to the movement bearing at $t-1$ and $\alpha_{1} \in (-\infty, \infty)$ is the bias coefficient in the direction $\psi_t$ (Fig. \ref{fig:vectors_sample}a). 

\subsubsection{Extending the BCRW to allow for menotaxis}
\begin{figure}  
    \centering
    \includegraphics[scale = 0.7]{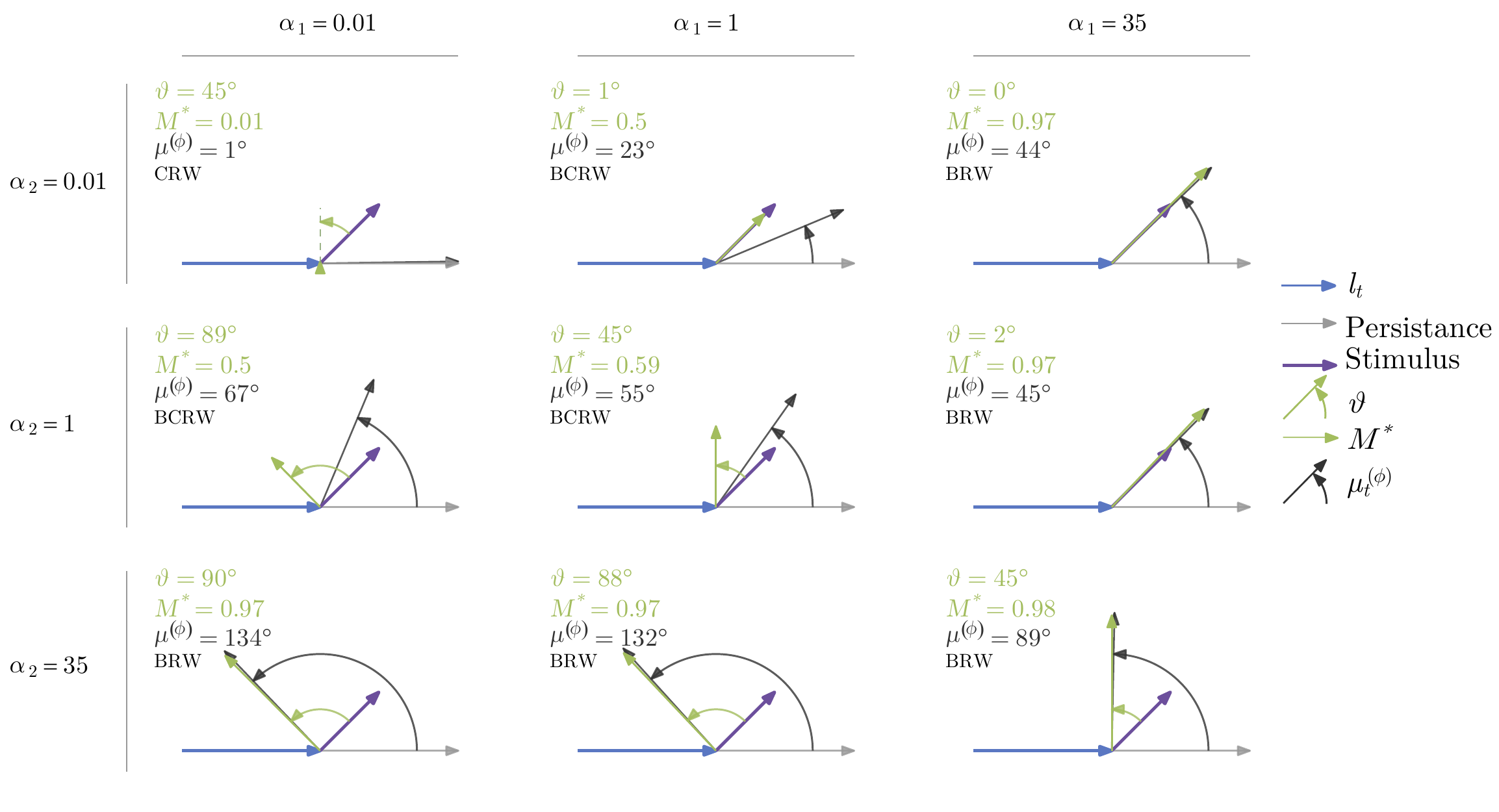}
    \caption{Illustration of various values of $\alpha_1$ and $\alpha_2$ and the corresponding direction of bias ($\vartheta$; green arc) relative to stimulus ($\psi_t = 45\degree$; purple), scaled magnitude of attraction ($M^\ast$; green vector). The expected turning angle ($\mu_t^{(\phi)}$; dark grey) is calculated as a trade-off between directional persistence (light grey) from the proceeding step (blue) and bias toward vector of attraction. For clarity, the dashed line in the first panel represents direction of $\vartheta$.}
     \label{fig:vectors}
\end{figure}

The BCRW described above can capture many behaviours exhibiting positive or negative taxis. However, many species exhibit menotactic movement where bias may be toward any angle relative to a stimulus (Fig. \ref{fig: biotaxis}). To capture behaviours with a bias toward an unknown angle of attraction $\vartheta$ relative to $\psi_t$, we propose modelling the mean turning angle as a trade-off between directional persistance, bias parallel to the direction of the stimulus, and bias perpendicular to the stimulus (Figs. \ref{fig:vectors_sample}a and \ref{fig:vectors}) following:
\begin{linenomath*}  \begin{align}
    \label{eq:BCRW.vt.TAm}
        \mu_{t}^{(\phi)} = \mathrm{atan2}(& \alpha_{1} \sin \psi_t + \alpha_{2} \cos \psi_t, \nonumber\\
                       & 1 + \alpha_{1} \cos \psi_t - \alpha_{2} \sin \psi_t),
\end{align}  \end{linenomath*}
where $\alpha_{1}$ represents the attraction coefficient parallel to $\psi_t$ as in Eq. \ref{eq:BCRW.TAm}, and $\alpha_{2} \in (-\infty, \infty)$ represents the bias coefficient toward $90\degree$ anti-clockwise of $\psi_t$. In this framework, the centre angle of attraction relative to a stimulus is represented by $\vartheta = \mathrm{atan2}(\alpha_{2}, \alpha_{1})$. Figs. \ref{fig:vectors_sample} and \ref{fig:vectors} depict how the mean turning angle is controlled by the two bias coefficients $\alpha_1$ and $\alpha_2$ and the angle of stimuli $\psi_t$. 

Given the non-linear circular-circular link function, the slope of Eq. \ref{eq:BCRW.vt.TAm} varies with $\psi_t$ and results in artefacts that may not be ecologically meaningful (e.g., a negative slope or slopes approaching $\infty$; see Appendix \ref{appendix:A}). We present an alternate formulation of $\mu_{t}^{(\phi)}$ with respect to $\psi_t$ with a constant slope in Appendix \ref{appendix:A}, however this cannot currently be implemented in user-friendly R packages such as \texttt{momentuHMM}.

We can represent where along the spectrum of CRW and BRW a behaviour lies with the magnitude of attraction $M = \sqrt{\alpha_{1}^2 + \alpha_{2}^2} \in [0, \infty)$. A value of $M = 0$ represents a CRW, values around 1 represent BCRWs with equal weight for directional persistence and bias toward $\vartheta$, and values approaching infinity represent BRWs toward $\vartheta$. For a more intuitive metric, we can scale $M$ to the unit interval using: 
\begin{linenomath*} \begin{equation}  
    \label{eq:TAm.bias}
        M^\ast = \frac{M}{1 + M}
\end{equation}  \end{linenomath*}
where $M^\ast \in [0,1)$ represents the scaled magnitude of attraction. The CRW and BRW are limiting cases of Eq. \ref{eq:TAm.bias}, where $M^\ast \rightarrow 0$ and $M^\ast \rightarrow 1$, respectively, while a BCRW would have an intermediate value of $M^\ast$. 

\subsubsection{Integrating multiple states using hidden Markov models}\label{Method:HMM}
Animal movement is behaviour-specific, and HMMs can be used to model telemetry data spanning multiple behaviours \citep{Langrock2012FlexibleExtensions, Patterson2008StatespaceMovement}. Using HMMs, we can combine multiple behaviours with distinct types of biased and correlated movement patterns. HMMs are defined by two components: an unobserved \textit{state process} (or \textit{hidden/latent process}) and an observed \textit{state-dependent process}. The state process assumes that animal behaviours are discrete latent states, $S_t \in \{1,...,N\}$, whose probabilities at any given time $t$ depend only on the state at the previous time step. That is, the state sequence $(S_1, ..., S_T)$ follows a Markov chain governed by state transition probabilities $\gamma_{i,j} = \mathrm{Pr}(S_{t+1} = j| S_{t} = i)$ for $i,j \in \{1,...,N\}$, which are summarized by the $N \times N$ transition probability matrix, $\mathbf{\Gamma} = (\gamma_{i,j})$. Second, the state-dependent process $\textbf{X} = (X_1,..., X_T)$ (where $X_t = \{l_t,\phi_t\}$) assumes that the probability of any given observation $X_t$, also known as the emission probability, depends only on the underlying latent state $S_t$ \citep{Langrock2012FlexibleExtensions, Patterson2008StatespaceMovement, Zucchini2016HiddenSeries}. That is, the step length and turning angle parameters and their respective coefficients (i.e., $\mu_t^{(l)}, \sigma_t^{(l)}, \mu_t^{(\phi)}, \kappa_t^{(\phi)}, \beta_1, \beta_2, \alpha_1, \alpha_2$) are assumed to be state-specific (i.e., $\mu_{S,t}^{(l)}, \sigma_{S,t}^{(l)}, ...$).  

For some behaviours, we may expect animals to have two different, but related, angles to stimuli. For example, during cross-wind movement, we may expect the animal to move $90 \degree$ to the left or right of the wind (i.e, ${\mu_{t}^{(\phi)}} = \psi_t \pm 90 \degree$). Such behaviours will have a bimodal distribution for their turning angle ${\mu_{t}^{(\phi)}}$ relative to stimulus $\psi_t$, which would be difficult to model with a single distribution. However, we can model such behaviour with two states: biased left of stimulus ($S^{(L)}$ with bias toward $\psi_t + \vartheta_{S^{(L)}}$) and biased right of stimulus  ($S^{(R)}$ with bias toward $\psi_t - \vartheta_{S^{(R)}}$). To reduce the number of estimated coefficients, these two states can share their parameters: $\kappa^{(\phi)}_{S^{(L,R)}}$, $\mu^{(l)}_{S^{(L,R)}}$, $\sigma^{(l)}_{S^{(L,R)}}$, and transition probabilities into and out of those states $\gamma_{i,j}$. If $\vartheta_{S^{(L)}}$ and $\vartheta_{S^{(R)}}$ are assumed to be symmetrical about $\psi_t$, $\alpha_{1,S}$ and $\alpha_{2,S}$ could be shared between symmetrical states by ensuring that $\alpha_{1, S^{(L)}} = \alpha_{1, S^{(R)}}$ and $\alpha_{2, S^{(L)}} = -\alpha_{2, S^{(R)}}$ in Eq. \ref{eq:BCRW.vt.TAm}.

\subsubsection{Accounting for mismatch in resolutions of data streams}\label{Method:accounting for mismatch}
The spatiotemporal resolution of data must be considered in research designs \citep{Adam2019JointModels,Bestley2013IntegrativePredator, Fagan2013SpatialMovement,  Mcclintock2020UncoveringModels}. If there is a significant mismatch in the resolutions among data streams, BCRW models may favour the higher resolution data streams, potentially at the expense of biological accuracy. Although the preferred direction of a BRW is primarily determined by external factors rather than the direction of the preceding step, the movement often appears correlated because the direction of bias is also highly correlated \citep[e.g., biased towards a distant target or toward temporally correlated stimulus; Fig. \ref{fig:vectors_sample}b;][]{Benhamou2006DetectingIndividual-dependent}. If persistence explains more of the observed orientation than environmental data, BCRW models may incorrectly classify BRWs as CRWs \citep{Benhamou2006DetectingIndividual-dependent, Codling2008RandomBiology}. This misclassification can occur if location resolution is very high or significantly higher than the environmental data, if environmental error is higher than the location data, or if the direction of stimulus remains relatively homogenous such that only the initial change in orientation exhibits taxis (Fig. \ref{fig:vectors_sample}b). Therefore, the first step in a BRW should be better explained by the environment than persistence even in the presence of error in environmental data (Fig. \ref{fig:vectors_sample}b).

To remedy the information loss associated with inadequate environmental data, we propose leveraging the information in the initial change in orientation by modelling the first step of a BRW state $B^{(f)}$ as a separate transition state from consecutive steps $B^{(c)}$. To ensure $B^{(f)}$ is modelled as a BRW, we must fix $\alpha_{1,B^{(f)}}$ or $\alpha_{2,B^{(f)}}$ to a large (positive or negative) value depending on the expected value of $\vartheta_{B^{(f)}}$. Second, to ensure that $B^{(f)}$ is fixed to just the first step in a BRW, we must ensure all states go through $B^{(f)}$ to get to $B^{(c)}$, that $B^{(f)}$ instantly transitions to $B^{(c)}$, and that $B^{(c)}$ cannot transition to $B^{(f)}$. These relationships can be enforced by ensuring the transition probability matrix $\mathbf{\Gamma}$ follows:
\begin{linenomath*} \begin{equation}  
\label{eq:general transition state TPM}
  \mathbf{\Gamma} = 
    \bbordermatrix{ 
               &  B^{(f)}           & B^{(c)}            &  3           & \ldots & N            \cr
      B^{(f)}  & 0                  & 1            &  0           & \ldots & 0            \cr
      B^{(c)}        & 0                  & \gamma_{B^{(c)},B^{(c)}} & \gamma_{B^{(c)},3} & \ldots & \gamma_{B^{(c)},N} \cr
      3        & \gamma_{3,B^{(f)}} & 0            & \gamma_{3,3} & \ldots & \gamma_{3,N} \cr
      \vdots & \vdots           & \vdots     & \vdots     & \ddots & \vdots     \cr
      N        & \gamma_{N,B^{(f)}} & 0            & \gamma_{N,3} & \ldots & \gamma_{N,N} \cr
      }. 
\end{equation}  \end{linenomath*}
To reduce the number of coefficients, the step length coefficients in Eq. \ref{eq:gamma_pdf} can be shared between $B^{(f)}$ and $B^{(c)}$. This method can be applied to any BRW state where the orientation of first step in the state is largely independent of the preceding step; including passive states driven by advection or strongly biased active states. 

\subsection{Simulation study} 
We conducted simulation studies to test two aspects of our proposed menotactic HMM. First, we validated the ability of our extended BCRW HMM to accurately recover states. Second, we investigated the effect of coarse environmental data on state detection and the ability of incorporating a one-step transitionary state to mediate the effects of environmental error. We used our polar bear case study as a base to develop the framework of our simulation studies.

\subsubsection{Polar bear ecology background} 
We investigated three key behaviours: a passive BRW drift state, a BCRW olfactory search state, and a CRW area restricted search state. The prime polar bear foraging habitat is the offshore pack ice during the winter months \citep[approximately January to May, depending on regional phenology;][]{Pilfold2012AgeSea, Stirling1975TheBehavior.}. The persistent motion of pack ice creates perpetual motion in the observed tracks, even when bears are stationary (e.g., still-hunting, prey handling, resting). One of the key drivers of sea ice motion is surface winds, whereby drift speed is approximately 2\% of that of wind speed and approximately -20\degree relative to the wind direction \citep{Tschudi2010TrackingIce}. Passive drift can be modelled as a BRW, with a predicted $\vartheta \approx \pm 20\degree$ (Fig. \ref{fig: biotaxis}d). 

In the spring, polar bears' primary prey, ringed seals (\textit{Pusa hispida}) occupy subnivean lairs for protection from predators and the environment \citep{Chambellant2012TemporalCanada, Florko2020DriversSea, Smith1975TheStructures}. Due to the large scale of the sea ice habitat and the cryptic nature of seals, polar bears rely heavily on olfaction to locate prey \citep{Stirling1978ComparativeAges, Togunov2017WindscapesCarnivore, Togunov2018Corrigendum:10.1038/srep46332}. The theoretical optimal search strategy for olfactory predators when they do not sense any prey is to travel cross-wind \citep{Baker2018AlgorithmsSpecies, Kennedy1974Pheromone-regulatedMoths,Nevitt2008EvidenceExulans.}, a pattern exhibited by polar bears \citep{Togunov2017WindscapesCarnivore,Togunov2018Corrigendum:10.1038/srep46332}. Olfactory search can be modelled as a BCRW, with a predicted $\vartheta \approx \pm 90\degree$ (Fig. \ref{fig: biotaxis}d). When in an area with suspected food availability, animals often exhibit decreased movement speed and increased sinuosity to increase foraging success \citep{Potts2014}, which has been documented in polar bears \citep{Auger-Methe2016EvaluatingGuilds}. Such area restricted search can be modelled as an unbiased CRW \citep[Fig. \ref{fig: biotaxis}d][]{Auger-Methe2015DifferentiatingWalk, Potts2014}. 

\subsubsection{Track simulation} \label{Method:sim}

We simulated movement tracks using a four-state HMM with drift $D$, olfactory search left of wind $O^{(L)}$, olfactory search right of wind $O^{(R)}$, and area restricted search $ARS$. The $D$ and $O^{(L,R)}$ states were biased in relation to uniquely simulated wind fields. Wind fields were generated in four steps: simulating pressure fields, deriving longitudinal and latitudinal pressure gradients, re-scaling the pressure gradients to represent wind velocities, and finally applying a Coriolis rotation to obtain final wind vectors (see Appendix \ref{appendix:B} for detail).

The $D$ state was defined as a passive BRW with mean step length determined by wind speed following Eq. \ref{eq:sl_m}, and turning angle following Eq. \ref{eq:TAm.bias} with bias toward $\vartheta_D = -15\degree$, as estimated from the polar bear case study below. $O^{(L)}$ and $O^{(R)}$ were defined as BCRWs following Eq. \ref{eq:TAm.bias} with biases toward $\vartheta_{O^{(L)}} = 90\degree$ and $\vartheta_{O^{(R)}} = -90\degree$ relative to wind and a large mean step length. Last, $ARS$ was defined as a CRW with no bias relative to wind and low mean step length. All coefficients used in the simulation are presented in Table \ref{tab:Sim_Sim_Par}. All tracks were simulated using the \texttt{momentuHMM} package \citep[see Appendix \ref{appendix:D} for details;][]{Mcclintock2018MomentuHMM:Movement}. 

\begin{table}[H]
  \caption{Coefficients used to simulate the four-state BCRW HMM movement track (left side) and the corresponding bias parameters (right side). Non-applicable values are indicated by `-'.} \label{tab:Sim_Sim_Par}
  \centering
  \begin{threeparttable}
  \begin{tabular}{@{}rcccccc||cc@{}} 
    \cmidrule[\heavyrulewidth](r){1-7} \cmidrule[\heavyrulewidth](l){8-9} 
    State      & $\beta_{1,S}$   & $\beta_{2,S}$  & $\sigma_S^{(l)}$ & $\alpha_{1,S}$ & $\alpha_{2,S}$   & $\kappa_S^{(\phi)}$ & $\vartheta_S (\degree)$ & $M_S^\ast$ \\    \cmidrule(r){1-7} \cmidrule(l){8-9} 
    $D$        & -2.2&0.1&-2.4&100&-26.8&5&-15&0.99\\ 
    $O^{(L)}$  & 0.1&    -            &-0.5&0&5&2&90&0.83 \\ 
    $O^{(R)}$  & 0.1&    -            &-0.5&0&5&2&-90&0.83 \\ 
    $ARS$      & -2.5&    -            &-2.8&0& - &0.5& - & 0  \\ \cmidrule[\heavyrulewidth](r){1-7} \cmidrule[\heavyrulewidth](l){8-9} 
    \end{tabular}
  \end{threeparttable}
\end{table}

In the first simulation study, movement relative to wind, $\psi_t$, was estimated from interpolated longitudinal and latitudinal pressure fields. In the second simulation study, the longitudinal and latitudinal wind fields were first spatially averaged to six resolutions (1, 2, 4, 8, 16, and 32 km) using the \texttt{raster} package \citep{Hijmans2016PackageModeling} then interpolated. For both simulation studies, we evaluated model accuracy by first predicting states from fit HMMs using the Viterbi algorithm \citep{Zucchini2016HiddenSeries}, merging analogous states (i.e., first drift $D^{(f)}$ and consecutive drift $D^{(c)}$; and $O^{(L)}$ and $O^{(R)}$), then calculating the proportion of correctly identified states (number of correctly identified steps / total number of steps). We also calculated recall (proportion of a simulated state correctly identified) for $D$ following $\sum$(correctly predicted $D$)/ $\sum$(simulated $D$).

\subsubsection{Fitting HMMs to simulations} \label{Method:sim.fit}

The first simulation validated the ability of our extended BCRW HMM to accurately recover states by comparing our model to a simpler CRW HMM. We fit 100 simulated tracks with two models: a three-state CRW HMM where mean drift speed was determined by wind but no states exhibit bias \citep[e.g.,][]{Ventura2020GadflyScales} and a four-state BCRW HMM with wind-driven drift speed and bias in two states.

As the CRW HMM assumes no bias is present, the mean turning angle was fixed to zero and not estimated for any state ($\mu_S^{(\phi)} = 0$). The BCRW HMM was formulated following the methods described in \ref{Method:sim}, with the $D$ and $O^{(L)}$ and $O^{(R)}$ states defined as biased relative to wind with both $\alpha_1$ and $\alpha_2$ being estimated. As we are primarily interested in differentiating the effect of incorporating menotaxis, the $D$ state step length was modelled as a function of wind speed as in Eq. \ref{eq:sl_m} in both CRW and BCRW HMMs.

 The second simulation study evaluated the effect of incorporating a transition state to alleviate information loss with lower environmental resolution. We down-scaled 100 simulations to six resolutions and fit two HMMs: the same four-state BCRW HMM used to simulate the tracks and a five-state HMM with drift divided into state for the first drift step $D^{(f)}$ and consecutive drift steps $D^{(c)}$ (hereafter, TBCRW HMM). Each simulated track was fit to the BCRW and TBCRW HMMs, which were compared by calculating the difference between their respective accuracies. 
 
 Drift was assumed to be a passive BRW relative to wind, which was ensured by fixing $\alpha_{1, D^{(f)}}$ to a large constant (in our case, $\alpha_{1, D^{(f)}} = 100$) as described in \ref{Method:accounting for mismatch}. $D^{(c)}$ was modelled as a biased state with respect to wind, however given the low resolution of wind data, no assumption was made on the direction of the bias relative to wind. Olfactory search was modelled as two states $O^{(L)}$ and $O^{(R)}$ that were biased with respect to wind, and we assumed these to be symmetrical about $0\degree$. Finally, we assumed $ARS$ to be a CRW independent of wind. These were modelled by modifying Eq. \ref{eq:BCRW.vt.TAm} to: 
\begin{linenomath*} \begin{equation} \label{eq:PB.BCRW.vt.TAm}
    \mu_{S,t}^{(\phi)} = 
    \begin{cases}
        \begin{aligned}
            &0               \\
            &\mathrm{atan2}(100\sin \psi_t + \alpha_{2,S} \cos \psi_t,\\
            &\quad \quad \quad 1 + 100\cos \psi_t - \alpha_{S,2} \sin \psi_t)\\
                        &\mathrm{atan2}(\alpha_{1,S} \sin \psi_t - \alpha_{2,S} \cos \psi_t,\\
            &\quad \quad \quad 1 + \alpha_{1,S} \cos \psi_t + \alpha_{2,S} \sin \psi_t)\\
            &\mathrm{atan2}(\alpha_{1,S} \sin \psi_t + \alpha_{2,S} \cos \psi_t,\\
            &\quad \quad \quad 1 + \alpha_{1,S} \cos \psi_t - \alpha_{2,S} \sin \psi_t)\\
        \end{aligned}
    \end{cases}
    \quad
    \begin{aligned}
    & \mathrm{if} \: S = ARS\\
    & \mathrm{if} \: S = D^{(f)}\\\\
    & \mathrm{if} \: S = O^{(R)}\\\\
    & \mathrm{Otherwise.}\\\\
    \end{aligned}
\end{equation}  \end{linenomath*}
To reduce the number of estimated coefficients, the coefficients for step length ($\beta^{(l)}_{1,S}$, $\beta^{(l)}_{2,S}$, and $\sigma^{(l)}_S$) were shared between $D^{(c)}$ and $D^{(f)}$ and between $O^{(L)}$ and $O^{(R)}$. For $O^{(L)}$ and $O^{(R)}$, $\kappa^{(\phi)}$, $\alpha_{1,S}$, and $\alpha_{2,S}$ were also shared along with the state transition probabilities $\gamma_{i,j}$. To ensure $D^{(f)}$ lasts one step, we fixed the transition probabilities following equation \ref{eq:general transition state TPM}. The transition probability matrix contained nine coefficients to be estimated following:
\begin{linenomath*} \begin{equation}  
\mathbf{\Gamma} = 
    \bbordermatrix{ 
          &  D^{(f)} & D^{(c)} & O^{(L)} & O^{(R)} & ARS   \cr
      D^{(f)} & 0 & 1 & 0 & 0 & 0 \cr
      D^{(c)}   & 0 & \gamma_{2,2} &\gamma_{2,3}  & \gamma_{2,3} & \gamma_{2,5} \cr
      O^{(L)} & \gamma_{3,1} & 0&\gamma_{3,3}&\gamma_{3,4}&\gamma_{3,5}\cr
      O^{(R)} & \gamma_{3,1} & 0&\gamma_{3,4}&\gamma_{3,3}&\gamma_{3,5}\cr
      ARS & \gamma_{5,1} & 0&\gamma_{5,3}&\gamma_{5,3}&\gamma_{5,5}\cr
}
\end{equation}  \end{linenomath*}

\subsection{Case study - Polar bear olfactory search and drift}

We illustrate the application of our HMM using global positioning system (GPS) tracking data from one adult female polar bear collared as part of long-term research on polar bear ecology \citep{Lunn2016DemographyBay}. The performed animal handling and tagging procedures were approved by the University of Alberta Animal Care and Use Committee for Biosciences and by the Environment Canada Prairie and Northern Region Animal Care Committee \citep{Stirling1989ImmobilizationArctic.}. We limited the analysis to one month during the peak foraging period between Apr 5, 2011 and May 5, 2011. The collar was programmed to obtain GPS locations at a 30 min frequency. To obtain the continuous environmental data necessary for HMMs, missing locations (n = 8; 0.56\%)  were imputed by fitting a continuous-time correlated random walk model using the \texttt{crawl} package in \texttt{R} \citep{Johnson2018Crawl:Data., Johnson2008Continuous-timeData, RCoreTeam2020R:Computing}. GPS locations were annotated with wind data using the ERA5 meteorological reanalysis project, which provides hourly global analysis fields at a 31 km resolution \citep{Hersbach2020TheReanalysis}. Wind estimates along the track were interpolated in space and time as described in \citet{Togunov2017WindscapesCarnivore, Togunov2018Corrigendum:10.1038/srep46332}. We modelled the polar bear track as a five-state TBCRW HMM as in the second simulation. All statistical analyses were done in \texttt{R} version 4.0.2 \citep{RCoreTeam2020R:Computing}. Reproducible code is presented in Appendix \ref{appendix:D}.

\section{Results}

\subsection{Simulation}
The first simulation served to validate that the proposed BCRW HMM can accurately recover states. As expected, the accuracy (mean and [2.5\%, 97.5\%] quantiles) of the BCRW HMM (0.96 [0.96, 0.96]) was higher than that of the unbiased CRW HMM (0.77 [0.77, 0.78]). In both models, olfactory search was the most accurately identified state (Table \ref{appendix:C}\ref{tab:Sim1_CM}). The BCRW HMM had higher precision in estimating all states compared to the CRW HMM. The CRW HMM had the lowest precision rate when predicting $D$, where it more frequently confused it for $ARS$ (Table \ref{appendix:C}\ref{tab:Sim1_CM}).

\begin{figure}
    \centering
    \includegraphics[scale = 1]{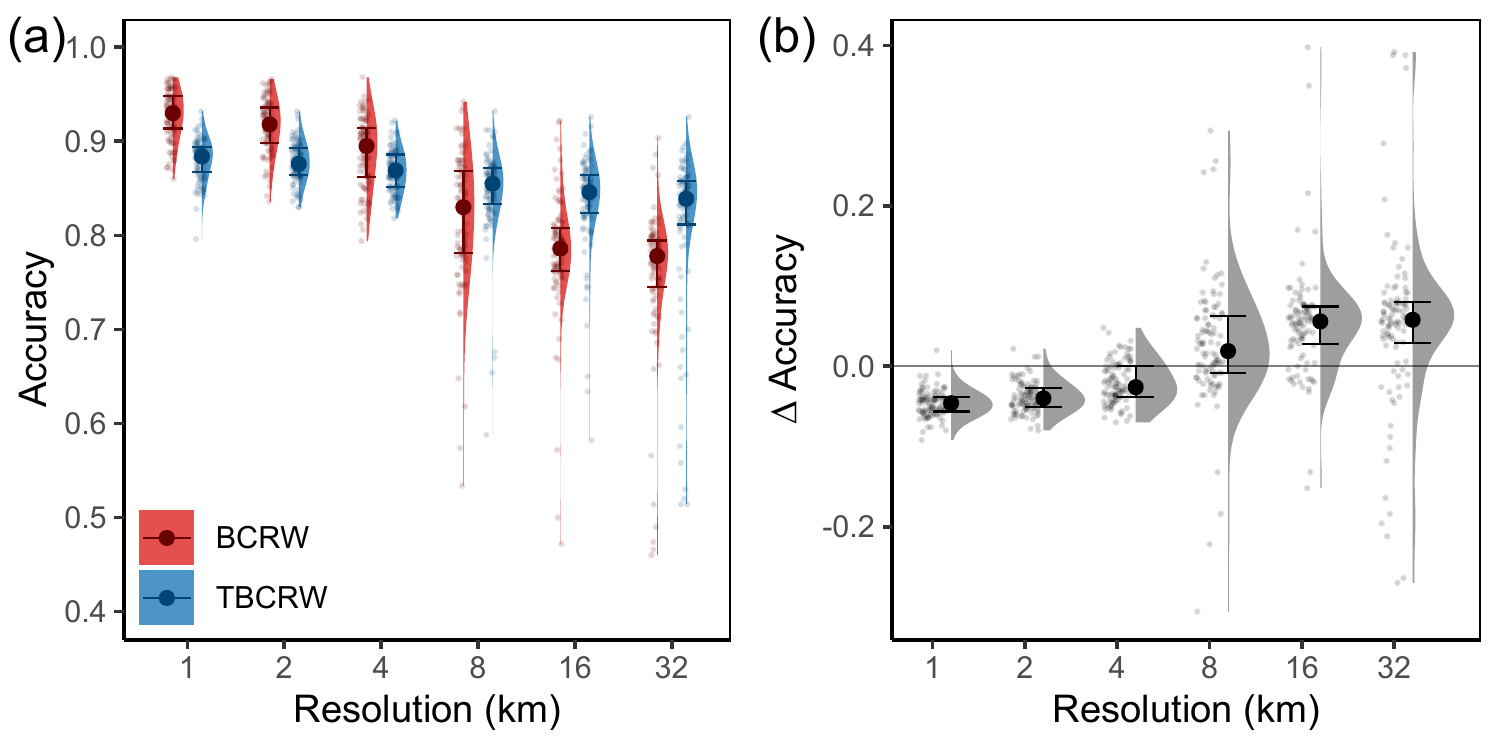}
    \caption{State prediction accuracy (a) and difference in accuracy (b) between the biased correlated random walk (BCRW) and transitionary-BCRW (TBCRW) HMMs ($\Delta$ accuracy = $\mathrm{accuracy_{TBCRW} - accuracy_{BCRW}}$). The accuracies of the 100 individual simulated tracks are represented by the transparent points, shaded areas represent the density across the 100 tracks, and opaque points represent the medians. Error bars span the 25\% and 75\% quantiles of the accuracies. 
    }
    \label{fig:Sim2_accuracy}
\end{figure}

When there was no environmental error, the basic BCRW HMM outperformed the TBCRW HMM (Fig. \ref{fig:Sim2_accuracy}). As the environmental resolution declined, the accuracy of both models also declined, however, the TBCRW HMM outperformed the BCRW HMM at resolutions coarser than 4 km (Fig. \ref{fig:Sim2_accuracy}). At finer resolutions, recall (mean and [25\%, 75\%] quantiles) of $D$ in the BCRW HMM was higher than in TBCRW HMM (e.g., $\mathrm{recall}_{BCRW,1 \mathrm km} = 0.88 [0.84, 0.91]$ vs $\mathrm{recall}_{TBCRW,1 \mathrm km}  = 0.74 [0.69, 0.77]$). At resolutions coarser than 8 km, recall of $D$ in the BCRW HMM was lower than in the TBCRW HMM (e.g., $\mathrm{recall}_{BCRW,32 \mathrm km} = 0.40 [0.31, 0.46]$ vs $\mathrm{recall}_{TBCRW,32 \mathrm km} = 0.61 [0.54, 0.67]$).

\begin{figure}
    \centering
    \includegraphics[scale = 1]{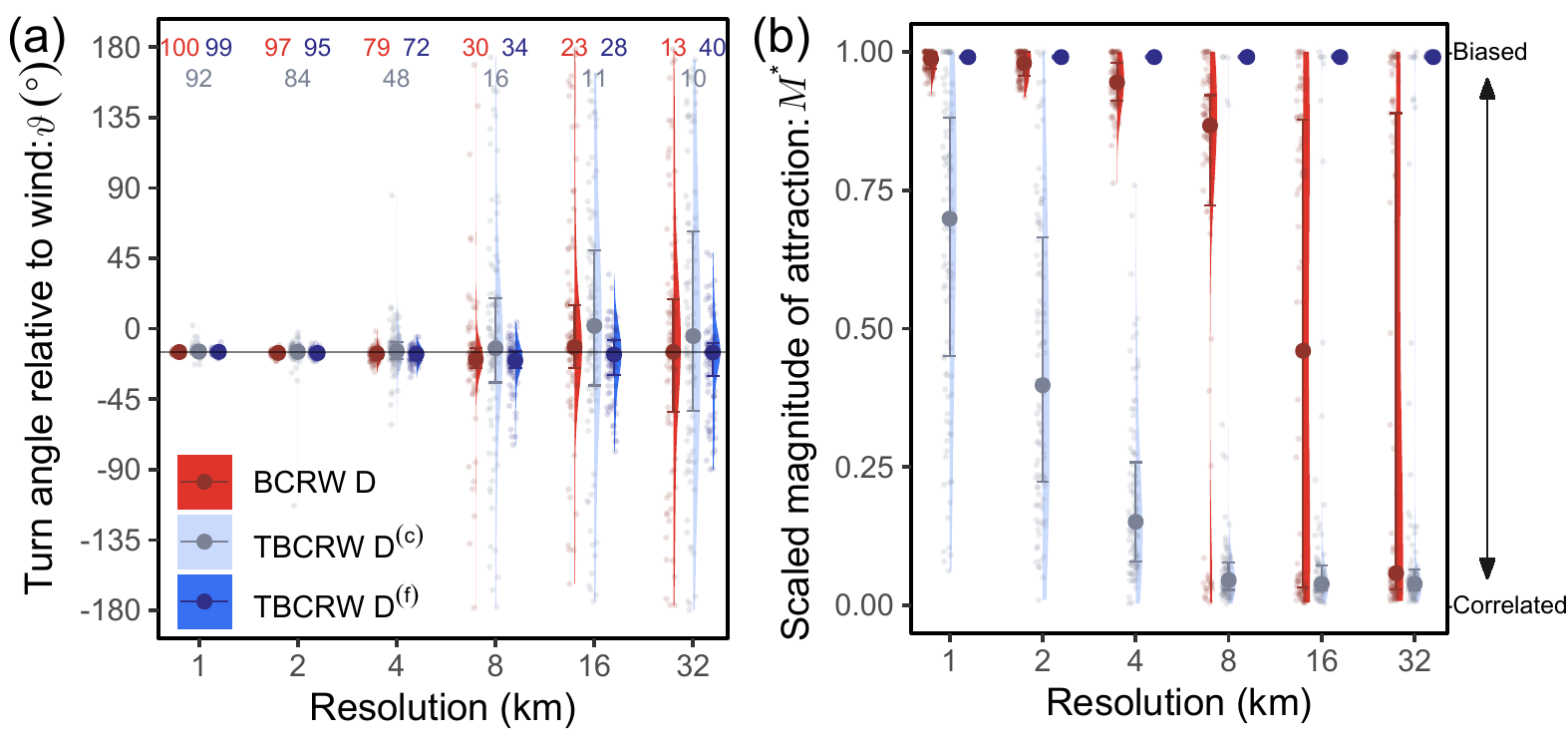}
    \caption{Estimated direction of bias relative to wind $\vartheta$ (a) and scaled magnitude of attraction $M^\ast$ (b) for the drift state ($D$) in the biased correlated random walk (BCRW) HMM (red), and first drift ($D^{(f)}$; blue) and consecutive drift ($D^{(c)}$; teal) states of the transitionary-BCRW (TBCRW) HMM. Values at the top of (a) represent what percent of the simulation runs estimated $\vartheta$ within $-15 \pm 5 \degree$ for each model. Points represent the raw estimates from the 100 simulated tracks, shaded areas represent the density across the 100 tracks, and opaque points represent the medians. Error bars span the 25\% and 75\% quantiles of the estimates. 
    }
    \label{fig:Sim2_bias}
\end{figure}

In both the BCRW HMM and TBCRW HMM, the mean estimates of $\vartheta_D$,  $\vartheta_{D^{(f)}}$, and $\vartheta_{D^{(c)}}$ across the 100 simulated tracks were close to the simulated $-15 \degree$, however, there was a marked difference in the spread of estimates across the simulations (Fig. \ref{fig:Sim2_bias}a). At resolutions 4 km or finer, the BCRW HMM provided reasonable estimates of $\vartheta_D$ and $M^\ast$ ($\geq 79\%$ of $\hat{\vartheta}_D \in \vartheta_D\pm 5 \degree$; Fig. \ref{fig:Sim2_bias}). At resolutions coarser than 4 km, the spread of estimated values increased with $\leq 30 \%$ of $\hat{\vartheta}_D \in \vartheta_D \pm 5 \degree$ (Fig. \ref{fig:Sim2_bias}a). According to the BCRW HMM, at a 16 km resolution, $D$ would be characterized as a BCRW ($M^\ast_{D} \approx 0.5$) and a BRW at 32 km ($M^\ast_{D,t} \approx 0$; Fig. \ref{fig:Sim2_bias}b). These patterns were exaggerated for the $D^{(c)}$ state in the TBCRW HMM, with the estimates of $\vartheta_{D^{(c)}}$ widening at 4 km resolutions and coarser and the rate of change $M^\ast_{D^{(c)}}$ suggested $D^{(c)}$ was a CRW (Fig. \ref{fig:Sim2_bias}). In contrast, the transitionary $D^{(f)}$ state in the TBCRW HMM was estimated more accurately than either $D^{(c)}$ or $D$, with more estimates of $\vartheta_{D^{(f)}}$ closer to the simulated $-15\degree$ at all resolutions (Fig. \ref{fig:Sim2_bias}a).


\subsection{Case Study: Polar bear olfactory search and drift}

\begin{figure}  
    \centering
    \includegraphics[scale=1.15]{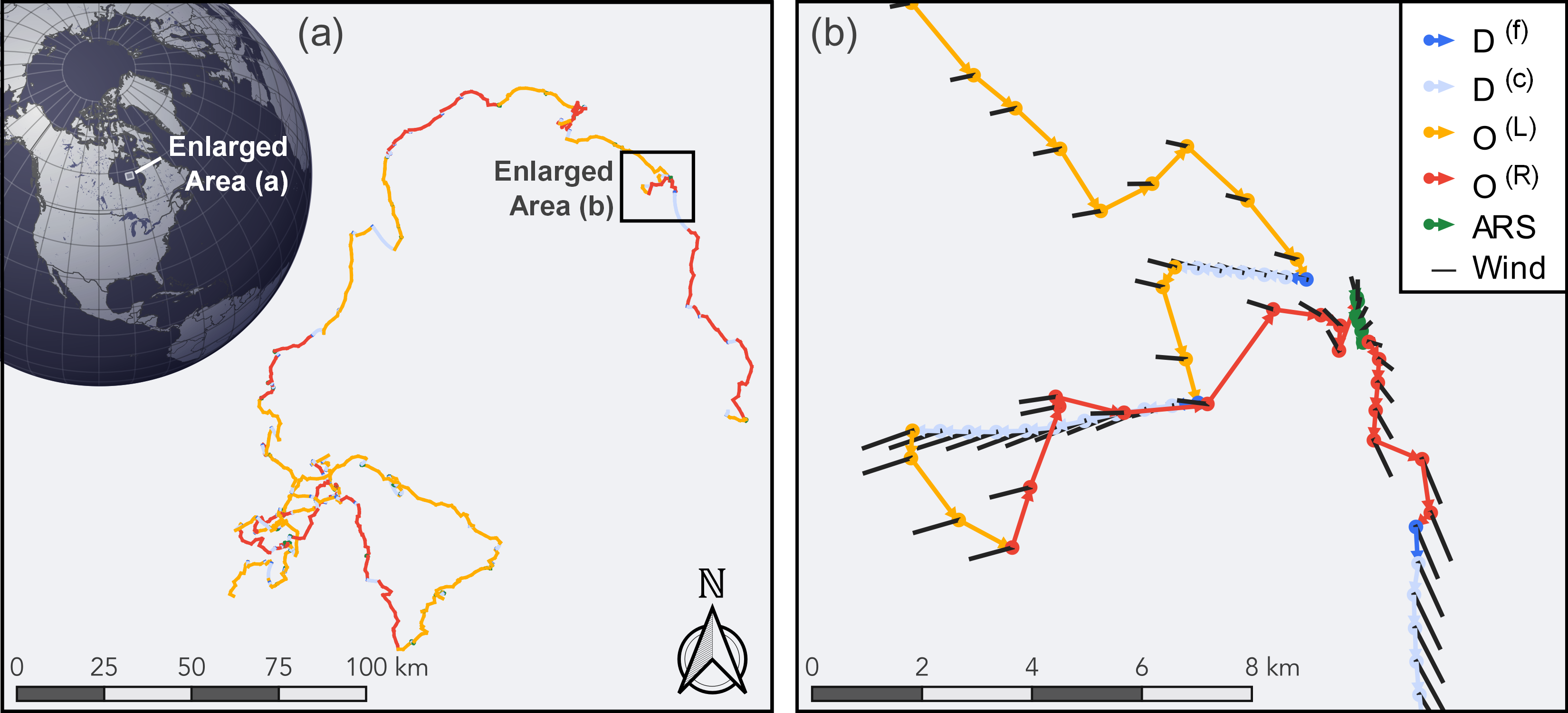}
    \caption{GPS track of a polar bear in spring 2011. Colours represent the predicted states of a five-state HMM determined using the Viterbi algorithm. Black lines represent direction and magnitude of wind.}
    \label{fig:PB_5B_map}
\end{figure}

The fitted HMM showed distinct movement characteristics among the five modelled states. Based on the stationary state distribution, the bear spent about 35\% of its time between $D^{(c)}$ and $D^{(f)}$, 47\% between $O^{(L)}$ and $O^{(R)}$, and 17\% in $ARS$. The only notably differences in transition probabilities were that $D^{(c)}$ was more likely to be followed by $O^{(L,R)}$ than $ARS$ and that the transition between $O^{(L)}$ and $O^{(R)}$ was significantly less likely than the probability of remaining within the same olfactory search state (Table \ref{tab:PB_HMM_TPM}). 

{\begin{table}[!h]
    \centering
    \caption{Estimated transition probability matrix, $\mathbf{\Gamma}$, for polar bear HMM. Values represent transition probabilities, $\gamma_{i,j} \pm \mathrm{SE}$ \citep[estimated using \texttt{momentuHMM};][]{Mcclintock2018MomentuHMM:Movement}, from state $i$ to $j$.}
    \label{tab:PB_HMM_TPM}
        \begin{equation*}
          \mathbf{\Gamma} =
            \bbordermatrix{ 
              & D^{(f)} & D^{(c)} & O^{(L)} & O^{(R)} & ARS   \cr 
              D^{(f)} & 0 & 1 & 0 & 0 & 0 \cr 
              D^{(c)}       & 0 & 0.83 \pm 0.02 & 0.07 \pm 0.01 & 0.07 \pm 0.01 & 0.02 \pm 0.01 \cr 
              O^{(L)} & 0.08 \pm 0.01 & 0 & 0.83 \pm 0.02 & 0.02 \pm 0.01 & 0.07 \pm 0.01 \cr 
              O^{(R)} & 0.08 \pm 0.01 & 0 & 0.02 \pm 0.01 & 0.83 \pm 0.02 & 0.07 \pm 0.01 \cr 
              ARS     & 0.07 \pm 0.02 & 0 & 0.08 \pm 0.01 & 0.08 \pm 0.01 & 0.77 \pm 0.03 \cr 
            } 
        \end{equation*}
\end{table} 
}

\begin{figure}[!h]  
    \centering
    \includegraphics[scale=0.8]{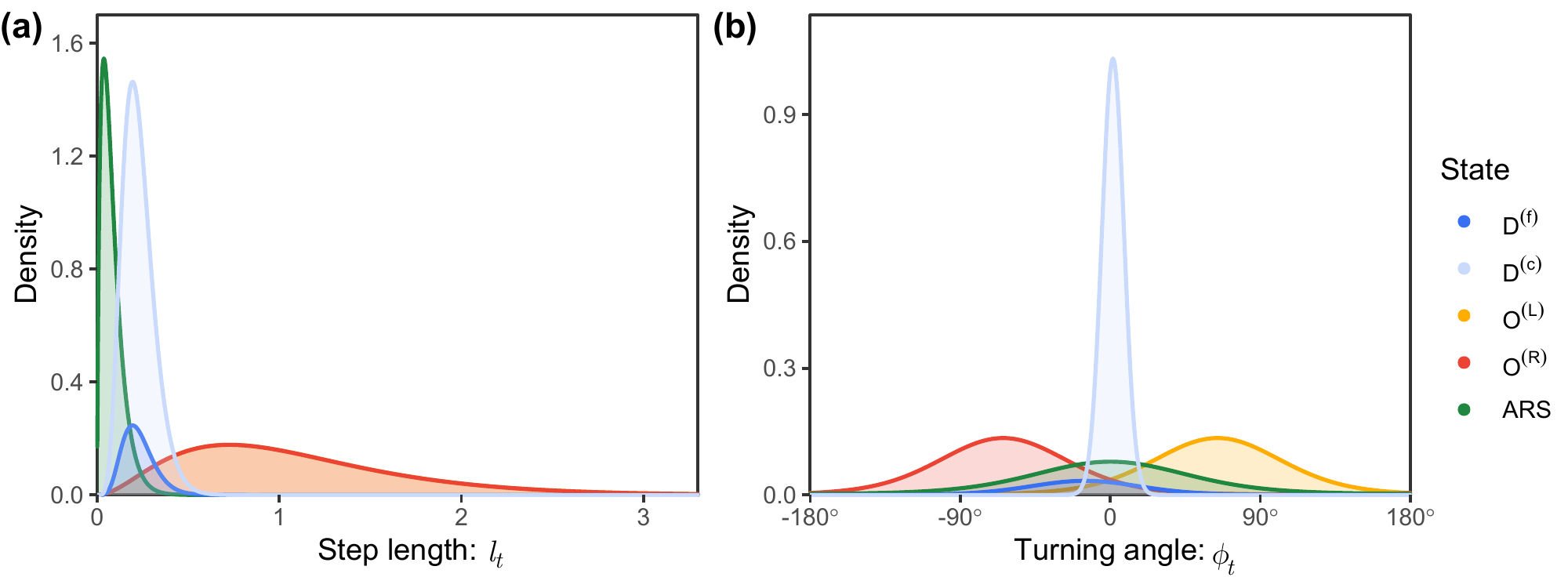}
    \caption{Probability density plots based on a fitted five-state HMM for (a) step length $l_t$ (km $30\mathrm{min}^{-1}$) assuming wind speed $r_t = 6\mathrm{m \, s}^{-1}$ and (b) turning angle $\phi_t$ (in radians) assuming relative angle $\psi_t = 0$. The height of each density distribution is scaled by the stationary state distribution of each respective state.}
    \label{fig:pb_SL_TA_PDF}
\end{figure}

\begin{figure}[!h] 
\centering
    \includegraphics[scale=0.8, width=18cm]{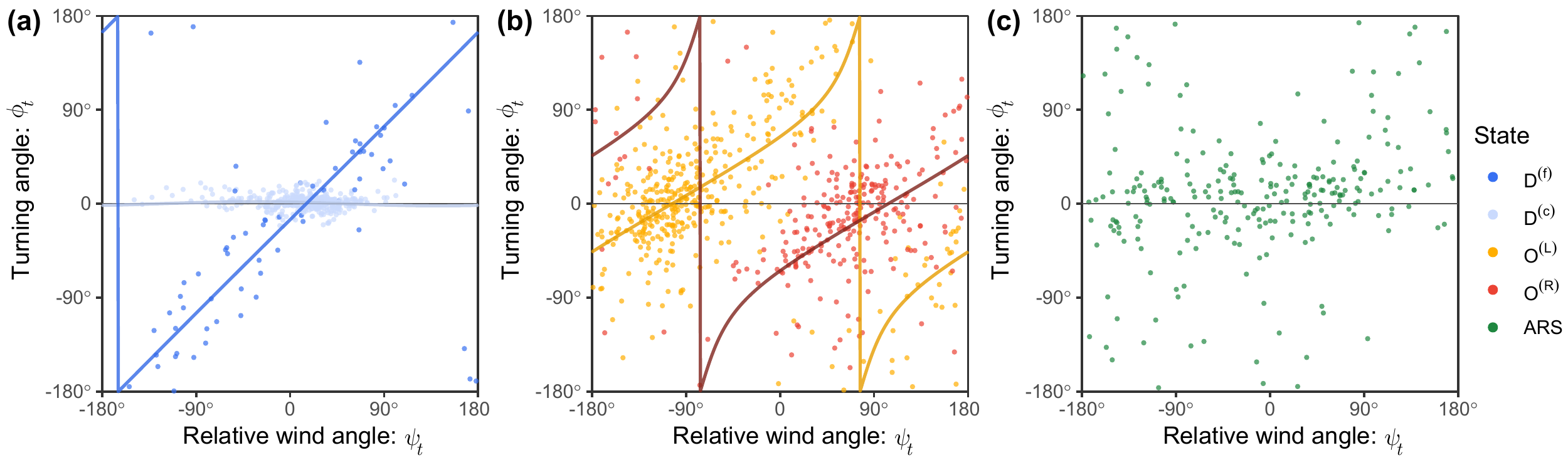}
    \caption{Turning angle $\phi_t$ (in radians) versus angle of wind relative to previous step $\psi_t$ (in radians). Lines represent the expected turning angle based on the estimated values of $\alpha_{1,S}$ and $\alpha_{2,S}$ given $\psi_t$.}
    \label{fig:pb_5B_TA_psi}
\end{figure}

First drift, $D^{(f)}$, was characterized as a passive BRW with mean direction and speed determined by wind. As described in \ref{Method:sim.fit}, the downwind bias coefficient was fixed to $\alpha_{1,D^{(f)}} = 100$ to ensure it was a BRW and crosswind bias coefficient was estimated at $\alpha_{2,D^{(f)}} = -27.65$, corresponding to an overall bias toward $\vartheta_{D^{(f)}} = -15\degree$ relative to wind (Figs. \ref{fig:pb_SL_TA_PDF}b and \ref{fig:pb_5B_TA_psi}a; Table \ref{tab:HB}). Turning angle concentration was moderate ($\kappa^{(\phi)}_{D^{(f)}} = 1.61$) and scaled magnitude of attraction was very high ($M^\ast_T = 0.99$), best characterizing $D^{(f)}$ as a BRW (Fig. \ref{fig:pb_5B_TA_psi}a; Table \ref{tab:HB}). 

\begin{table}[!h]
\centering
\caption{Estimated parameters and coefficients for five-state polar bear HMM.}
\label{tab:HB}
\begin{tabular}{lllll}
\toprule
Data stream & Parameter & Coefficient & States & Estimate (SE) \\ \hline
 & \multirow{4}{*}{$\mu^{(l)}$} & $\beta_1$ & $D^{(f)}$, $D^{(c)}$ & -2.118 (0.042) \\
\multirow{6}{*}{Step length} &  & $\beta_2$ & $D^{(f)}$, $D^{(c)}$ & 0.107 (0.004) \\
 &  & $\beta_1$ & $O^{(L)}$, $O^{(R)}$ & 0.087 (0.027) \\
 &  & $\beta_1$ & $ARS$ & -2.501 (0.106) \\ \cline{2-5} 
 & \multirow{3}{*}{$\sigma^{(l)}$} &  & $D^{(f)}$, $D^{(c)}$ & -2.445 (0.037) \\
 &  &  & $O^{(L)}$, $O^{(R)}$ & -0.462 (0.044) \\
 &  &  & $ARS$ & -2.804 (0.153) \\ \hline
 & \multirow{7}{*}{$\mu^{(\phi)}$} & $\alpha_1$ & $D^{(f)}$ & 100 (0) \\
 &  & $\alpha_2$ & $D^{(f)}$ & -27.645 (8.409) \\
 &  & $\alpha_1$ & $D^{(c)}$ & -0.020 (0.015) \\
\multirow{8}{*}{Turning angle} &  & $\alpha_2$ & $D^{(c)}$ & 0.027 (0.008) \\
 &  & $\alpha_1$ & $O^{(L)}$, $O^{(R)}$ & -0.332 (0.062) \\
 &  & $\alpha_2$ & $O^{(L)}$, $O^{(R)}$ & 1.385 (0.125) \\
 &  & $\mu^{(\phi)}$ & $ARS$ & 0 (0) \\ \cline{2-5} 
 & \multirow{4}{*}{$\kappa^{(\phi)}$} &  & $D^{(f)}$ & 1.131 (0.158) \\
 &  &  & $D^{(c)}$ & 4.325 (0.112) \\
 &  &  & $O^{(L)}$, $O^{(R)}$ & 0.843 (0.068) \\
 &  &  & $ARS$ & 0.476 (0.138) \\ \bottomrule
\end{tabular}
\end{table}

\begin{figure}[!h]  
    \centering
    \includegraphics[scale=0.8]{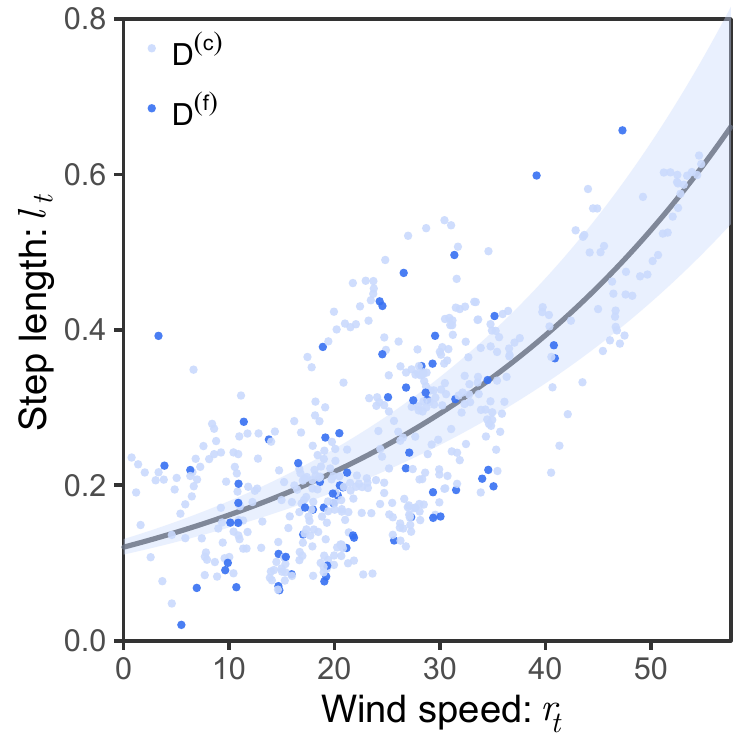}
    \caption{Curve representing the estimated mean step length parameter $\mu^{(l)}_t$ (in km 30m$^{-1}$) as a function of wind speed $r_t$ (in km h$^{-1}$). Shaded area represents 95\% confidence interval. Points represent observed step lengths for first drift ($D^{(f)}$) and consecutive drift ($D^{(c)}$) states.}
    \label{fig:PB_SL_WndSpd}
\end{figure}

The mean observed direction of drift relative to wind was $13\degree$ (estimated from a von Mises distribution fit to the direction of wind for steps classified as $D^{(f)}$ or $D^{(c)}$). However, turning angle during $D^{(c)}$ showed almost no bias relative to wind ($\alpha_{1,D^{(c)}} = -0.02$ and $\alpha_{2,D^{(c)}} = 0.03$; $M^\ast_{D^{(c)}} = 0.03$) and turning angle concentration was very high ($\kappa^{(\phi)}_{D^{(c)}} = 75.60$), thus best characterizing $D^{(c)}$ as a CRW with high persistence (Figs. \ref{fig:pb_SL_TA_PDF}b and \ref{fig:pb_5B_TA_psi}a; Table \ref{tab:HB}). These results mirror those observed in the second simulation, which characterized $D^{(f)}$ as a BRW and $D^{(c)}$ as a CRW (Fig. \ref{fig:Sim2_bias}). The step length of $D^{(f)}$ and $D^{(c)}$ were characterized by a small step length that was explained by an exponential relationship to wind speed (Fig. \ref{fig:PB_SL_WndSpd}). At the median wind speed of 21.0 km h$^{-1}$, the mean step length of $D^{(f,c)}$ was $0.22 \pm 0.09$ km 30m$^{-1}$ ($\mu^{(l)}_{D^{(f,c)}} \pm \sigma^{(l)}_{D^{(f,c)}}$), or $\approx2\%$ of wind speed (Fig. \ref{fig:pb_SL_TA_PDF}a; Table \ref{tab:HB}).

Olfactory search was characterized as a fast BCRW relative to wind. The estimated mean step length was $1.09 \pm 0.63$ km ($\mu^{(l)}_{O^{(L)},O^{(R)}} \pm \sigma^{(l)}_{O^{(L)},O^{(R)}}$; Fig. \ref{fig:pb_SL_TA_PDF}a; Table \ref{tab:HB}). Bias downwind during olfactory search was $\alpha_{1,O^{(L)},O^{(R)}} = -0.332$ and bias crosswind was $\alpha_{2,O^{(L)},O^{(R)}} = \pm 1.385$, corresponding to an overall bias toward $\vartheta_{O^{(L)},O^{(R)}} = \pm 103 \degree$ relative to wind (Figs. \ref{fig:pb_SL_TA_PDF}b and \ref{fig:pb_5B_TA_psi}b; Table \ref{tab:HB}). The turning angle concentration was moderate ($\kappa^{(\phi)}_{O^{(L)},O^{(R)}} = 2.324$) as was the scaled magnitude of attraction ($M^\ast_{O^{(L)},O^{(R)}} = 0.588$), best characterizing olfactory search as a BCRW (Fig. \ref{fig:pb_5B_TA_psi}b; Table \ref{tab:HB}).

ARS was characterized as a slow CRW with no bias relative to wind. The estimated mean step length was $0.08 \pm 0.06$ km ($\mu^{(l)}_{ARS} \pm \sigma^{(l)}_{ARS}$; Fig. \ref{fig:pb_SL_TA_PDF}a; Table \ref{tab:HB}). Mean turning angle was fixed to zero and the turning angle concentration was the lowest among the states ($\kappa^{(\phi)}_{ARS} = 1.61$), best characterizing ARS as a CRW with low persistence (Figs. \ref{fig:pb_SL_TA_PDF}b and \ref{fig:pb_5B_TA_psi}c; Table \ref{tab:HB}).

\section{Discussion}
Behaviours with biased movement are common among animals for obtaining resources and avoid costs \citep{Bailey2018NavigationalMovement, Michelot2017EstimationPredators,  Ylitalo2020AnalysisModels}. Here, we described two extensions to HMMs to identify and characterize menotaxic behaviours and BRWs. By modelling turning angle bias with a component parallel to and a component perpendicular to stimuli, menotexic behaviours with bias toward any angle can be modelled. Second, we outline the use of a one-step `transitionary' state for taxic BRWs, which can improve the accuracy of state detection and estimation of the direction of bias when the resolution of the environmental data is coarse relative to the animal track. We illustrated the application of these methods for detecting olfactory search and passive drift from both simulated data and polar bear tracking data using the readily accessible and well documented \texttt{momentuHMM} package in R \citep{Mcclintock2018MomentuHMM:Movement}. To further aid in the implementation of these methods, we have provided a tutorial with reproducible R code in Appendix \ref{appendix:D}. 

Given the ubiquity of taxes exhibited among animals, other studies have integrated bias into their movement models. However, these have exclusively been simple attractive or repulsive bias \citep[i.e., positive and negative taxis; e.g.,][]{Benhamou2014OfMovements, Mcclintock2018MomentuHMM:Movement, Michelot2017EstimationPredators}. Further, although there has been much investigation into mechanisms and consequences of behaviours with nonparallel bias relative to the direction of stimuli, such menotaxis has yet to be mechanistically integrated in a movement model. Typically, menotaxis has been studied independently of the movement process or using \textit{post hoc} analysis and is identified either using visual assessment or basic descriptive statistics  \citep[e.g.,][]{Mestre2014, Paiva2010FlightEnvironment, Togunov2017WindscapesCarnivore, Togunov2018Corrigendum:10.1038/srep46332, Ventura2020GadflyScales}. Such methods are appropriate for some analyses, however, they may prohibit investigating more nuanced relationships between animals, their behaviour, and the environment. The direction of taxis may be unknown, incorrectly assumed \textit{a priori}, or may be affected by other factors (e.g., environmental covariates, internal state). Not accounting for these interactions may lead to incorrectly classifying movement or incorrectly estimating the direction or strength of bias. For instance, conventional methods for analysis of tracking data in mobile environments account for involuntary motion by simply subtracting the component of drift from the movement track \citep[e.g.,][]{Blanchet2020Space-useBears, Gaspar2006MarineTrack, Klappstein2020PatternsBay, Safi2013FlyingFlight}. However, this type of correction does not account for the error often in the environmental data \citep{Dohan2010MonitoringSensors, Togunov2020OpportunisticCanada, Yonehara2016FlightDirection}. Our methods overcome some of these limitations by building on the robust framework of HMMs, which are  relatively flexible to uncertainty in both the track and environmental data by distinguishing between latent state and state-dependent processes \citep{Mcclintock2012AWalks, Zucchini2016HiddenSeries}. 

Our first simulation study validated the ability of our proposed BCRW HMM (i.e., integrating anemotaxis into predicting the turning angle) to accurately identify states. Compared to an unbiased HMM, our method more reliably identified all three behavioural states, with a marked increase in accuracy (60 percentage points) for the drift state compared to more traditional HMM models. In cases where the speed of movement (i.e., step lengths) among different behaviours become more similar, a conventional HMM using only step length and turning angle may be completely unable to differentiate states, while our model may still be able to differentiate taxes.

When modelling measured data, the estimated parameters reflect both the underlying process of interest (e.g., a connection between movement, wind, and drift) as well as any underlying error in the data \citep{Bestley2013IntegrativePredator}. For instance, a low coefficient for a covariate may correctly reflect a weak ecological relationship or be an incorrect artefact of data with high error \citep{Bestley2013IntegrativePredator}. Our second simulation study demonstrated that at coarser resolutions of environmental data, a simpler BCRW HMM was prone to incorrectly characterizing BRWs as CRWs. We showed that the use of a one-step transitionary state can help recover some of the information lost for taxic BRWs in the presence of environmental error. The use of a transitionary state yielded two advantages: reduced misclassification of BRWs and improved estimation of the direction of bias. Although the use of a transitionary drift state caused consecutive drift steps to be misclassified as a CRW, the transitionary state was able to recover information on bias that would otherwise be lost entirely using conventional models. Employing a transitionary state for BRWs has one important caveat: because the transitionary state lasts for precisely one step, the model requires there to be a sufficient number of state transitions to the BRW for sufficient power to accurately estimate bias coefficients. To obtain a sufficient number of BRW transitions may require longer tracking or sharing coefficients among different animals. The utility of employing a transitionary state method depends on the system, but generally, it may be advantageous if the temporal resolution of the environmental data is coarser than the tracking data, if the angular error in the environmental data tends to be greater than the turning angle, or if high homogeneity in the environmental data carries error across multiple steps. In our case, the use of a transitionary state was beneficial as soon as the track data resolution was equal to or higher than the temporal resolution of the wind data (data not shown).

Different behaviours have unique fitness consequences and unique relationships with the environment, thus identifying the underlying behavioural context is critical to effectively interpret observed data \citep{Roever2014TheSelection, Wilson2012BeyondSelection}. This study was the first to identify stationary behaviour in polar bear tracking data, which made up a notable 35\% of the track duration. The drifting state encapsulates several distinct behaviours including rest, sheltering during adverse conditions, still hunting by a seal breathing hole, or prey handling \citep{Stirling1974MidsummerMaritimus, Stirling2016BehaviorMaritimus}. To distinguish between this mixture of behaviours we may investigate the effect of time or environmental conditions on the transition probabilities between states (i.e., relaxing the model assumption of homogenous state transition probability matrix). In some cases, the strength or direction of bias may depend on other factors. For example, bias to the centre of an animal's home range may depend on its distance from that centre \citep{Mcclintock2012AWalks}, or the strength of bias relative to wind during passive advection being proportional to wind speed \citep[e.g.,][]{Yu2020EvaluationHIRHAM-NAOSIM}. Such interactions with the bias can be accomplished by modelling bias coefficients, $\alpha_1$ and $\alpha_2$, as functions of covariates (e.g., distance to target, or magnitude of stimuli). Another important extension is to model multiple biases simultaneously, as animal movement is often driven by several competing goals \citep[e.g., navigation in flocking birds;][]{Nagy2010HierarchicalFlocks}. These extensions can all be readily accommodated using \texttt{momentuHMM} \citep{Mcclintock2018MomentuHMM:Movement}. In cases where simultaneous biases interact in complex ways, more advanced extensions may be required \citep{Mouritsen2018Long-distanceAnimals}. In mobile environments, such as birds in flight, the movement of the animal itself contains information on the flow since it is influenced by the currents and advection may affect the appearance of movement in all behaviours \citep{Goto2017AsymmetryOcean, Wilmers2015TheEcology, Yonehara2016FlightDirection}. If the magnitude of advection is comparable to the speed of voluntary movement, explicitly accounting for advection in all behaviours becomes increasingly important \citep{Auger-Methe2016HomeIce, Gaspar2006MarineTrack, Yonehara2016FlightDirection}. Our model can serve as a starting point for modelling menotactic BCRWs without making assumptions on the strength or direction of bias or assuming error-free movement and environmental data. The methods we present in this paper are simple extensions to conventional movement models. They can be readily applied to animal tracking data to characterise menotactic behaviours and open new avenues to investigate more nuanced and mechanistic relationships between animals and their environment. 

\section{Acknowledgments}
Financial and logistical support of this study was provided by the Canadian Association of Zoos and Aquariums, the Canadian Research Chairs Program, the Churchill Northern Studies Centre, Canadian Wildlife Federation, Care for the Wild International, Earth Rangers Foundation, Environment and Climate Change Canada, Hauser Bears, the Isdell Family Foundation, Kansas City Zoo, Manitoba Sustainable Development, Natural Sciences and Engineering Research Council of Canada, Parks Canada Agency, Pittsburgh Zoo Conservation Fund, Polar Bears International, Quark Expeditions, Schad Foundation, Sigmund Soudack and Associates Inc., Wildlife Media Inc., and World Wildlife Fund Canada. We thank B.T. McClintock for assistance with momentuHMM and E. Sidrow for reviewing the manuscript.

\section{Authors' contributions}
RRT and MAM conceived the ideas and designed methodology; NJL and AED conducted fieldwork; RRT conducted the analyses and prepared the manuscript. All authors contributed critically to the drafts and gave final approval for publication.

\section{Data Availability}
The location data of the polar bear case study is available on UAL Dataverse \citep{Derocher2021}. Code to reproduce the simulations and the case study is available on Zenodo \citep{Togunov2021Replicationmodels}. Wind data were obtained from the ERA5 meteorological reanalysis project \citep{Hersbach2020TheReanalysis}.


\appendix

\renewcommand\thefigure{\thesection.\arabic{figure}}    
\section{Appendix: Turning Angle Link Function} \label{appendix:A}
\setcounter{figure}{0}   

\begin{figure}  
    \centering
    \includegraphics[scale = 0.5]{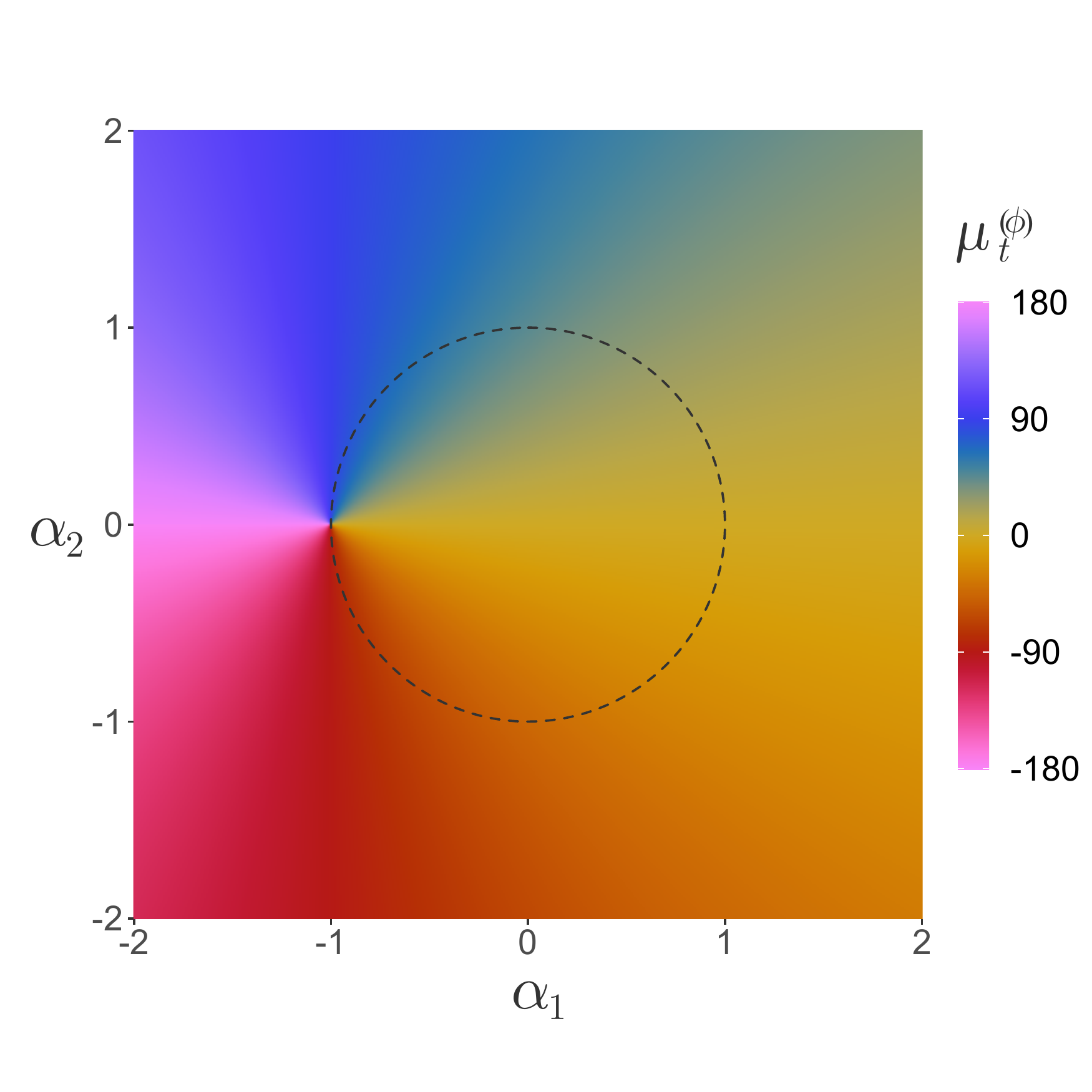}
    \caption{The expected turning angle $\mu^{(\phi)}_t$ in degrees given values of $\alpha_1$ and $\alpha_2$, assuming movement is in the same direction as stimulus (i.e., $\psi_t = 0 \degree$). As $\psi_t$ increases, this plot would effectively rotate clockwise about $(\alpha_1 = 0, \alpha_2 = 0)$ by the value of $\psi_t$. When $\sqrt{(\alpha_1^2 + \alpha_2^2)} \geq 1$ (on or outside the circle), increasing increasing $\psi_t$ would always result in increasing turning angle. However, when $\sqrt{(\alpha_1^2 + \alpha_2^2)} < 1$ (inside the circle) and as $|\psi_t| \rightarrow 180\degree$, increasing $\psi_t$ can result in decreasing turning angle. 
    }
     \label{fig:alph_vt}
\end{figure}

\begin{figure}[H]
    \centering
    \includegraphics{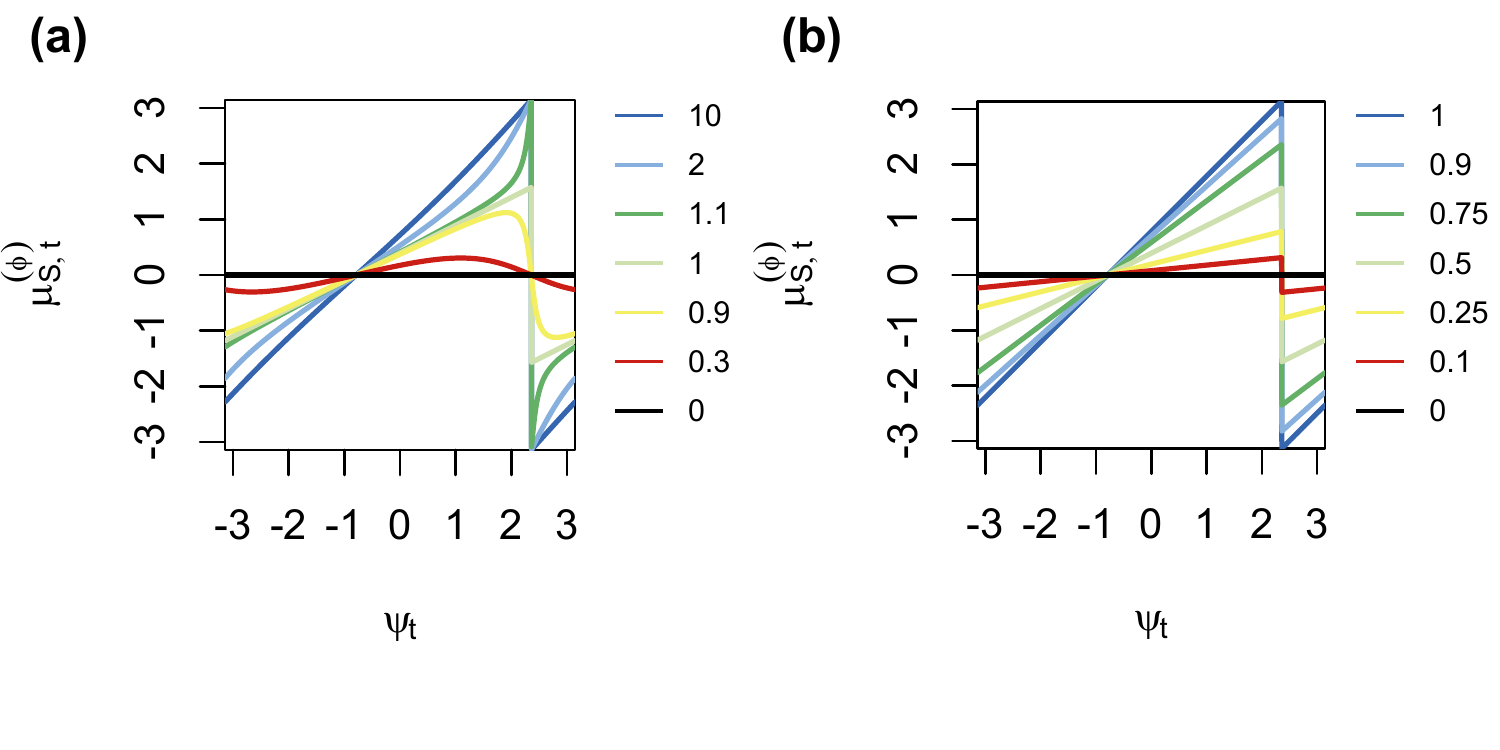}
    \caption{Predicted turning angle ${\mu_{S,t}^{(\phi)}}$ relative to direction of stimulus $\psi_t$ assuming bias is toward $\vartheta = 45\degree$. Panel `a' represents the predicted turning angle using the circular-circular von Mises distribution of \citet{Rivest2016AAnalysis}. The $\alpha_{1,S}$ and $\alpha_{2,S}$ are constraint such that magnitude of attraction $M_S = \sqrt{\alpha_{1,S}^2 + \alpha_{2,S}^2}$ is between 0 (black) and 10 (dark blue) as shown. Panel `b' represents the predicted turning angle using our proposed circular-linear von Mises distribution defined in Eq. \ref{eq:BCRW.TAm.alt} with a attraction coefficient $\lambda_S$ between 0 (black) and 1 (dark blue) as shown. Both axes are in radians.}
    \label{fig:TAm_psi_CC_CL}
\end{figure}

In this paper, we modelled mean turning angle ${\mu_{S,t}^{(\phi)}}$ relative to external stimuli $\psi_t$ using the circular-circular link function of \citet{Rivest2016AAnalysis}. By modelling movement as a bias in the same direction as a stimulus $\alpha_1$ and a bias perpendicular to a stimulus $\alpha_2$, we can predict movement with a bias toward any angle relative to the stimulus Fig. \ref{fig:alph_vt}. However, the circular-circular (i.e., non-linear) link function produces artefacts that may not have ecological significance. First, we define the rate of change of ${\mu_{S,t}^{(\phi)}}$ with respect to $\psi_t$ (i.e., the turning angle rate of change) as the first-order derivative of Eq. \ref{eq:BCRW.vt.TAm}
\begin{linenomath*} \begin{equation}  
    \label{eq:TAm.bias_roc}
        {\mu_{S,t}^{(\phi)}}' = \frac{\alpha_{1,S}^2 + \alpha_{1,S}\cos(\psi_t) + \alpha_{2,S}^2 - \alpha_{2,S} \sin(\psi_t)}{\alpha_{1,S}^2 + 2\alpha_{1,S}\cos(\psi_t) + \alpha_{2,S}^2 - 2\alpha_{2,S}\sin(\psi_t) + 1}.
\end{equation}  \end{linenomath*} 
We might expect ${\mu_{S,t}^{(\phi)}}'$ to be monotonically positive. For instance, as the direction of external stimuli rotate anti-clockwise, so should the predicted turning angle. The circular-circular distribution can produce negative values of ${\mu_{S,t}^{(\phi)}}'$ occur when the the movement of a behaviour is closer to a CRW (i.e., magnitude of bias $\sqrt{\alpha_{1,S}^2+\alpha_{2,S}^2} < 1$) and the movement is opposite to the direction of bias (i.e., $\psi_t \rightarrow -\vartheta + 180 \degree$; red and yellow curves in Fig. \ref{fig:TAm_psi_CC_CL} a). Second, values of ${\mu_{S,t}^{(\phi)}}' \rightarrow \infty$ as $\sqrt{\alpha_{1,S}^2+\alpha_{2,S}^2} \rightarrow 1^+$; this occurs over a narrow interval of values of $\alpha_{1,S}$ and $\alpha_{2,S}$ and when movement is opposite the direction of bias (i.e., $\psi_t \rightarrow \vartheta + 180 \degree$; teal curve in Fig. \ref{fig:TAm_psi_CC_CL} a).

An alternative way to formulate biased behaviours is to model $\mu_{S,t}^{(\phi)}$ explicitly relative to a the direction of bias $\vartheta_S$ and the strength of bias. Specifically, assuming a von Mises distribution for turning angle, the predicted mean turning angle can be modelled by
\begin{linenomath*} \begin{equation}  
    \label{eq:BCRW.TAm.alt}
    \mu_{S,t}^{(\phi)} = \arctan\Big(\tan\frac{\psi_t + \vartheta_{S}}{2}\Big)2\lambda_{S},
\end{equation}  \end{linenomath*}
where $\lambda_S \in [0,1]$ represents the state-specific bias coefficient trade-off between directional persistence and bias toward the expected state-specific orientation relative to stimulus $\vartheta_{S} \in (-\pi, \pi]$, and $\psi_t$ is the observed angle of stimulus relative to the movement bearing at time $t-1$ (Fig. \ref{fig:vectors}). The tan and arctan arguments ensure that the predicted values of $\mu_{S,t}^\phi$ fall within $(-\pi, \pi]$ and that the functions are linear, and $\lambda_S$ is effectively the slope coefficient determining the strength of bias. Specifically, $\lambda_{S} = 0$ represents a CRW (black in Fig. \ref{fig:TAm_psi_CC_CL} b), $\lambda_{S} = 1$ represents a BRW toward $\vartheta_S$ (purple in Fig. \ref{fig:TAm_psi_CC_CL} b), and $\lambda_{S} = 0.5$ represents a BCRW with equal weight to directional persistence and bias toward $\vartheta_S$ (green in Fig. \ref{fig:TAm_psi_CC_CL} b). This formulation directly models the parameters of interest and does not exhibit nonlinear artefacts as in the circular-circular distribution (Fig. \ref{fig:TAm_psi_CC_CL}).

We developed our BCRW HMM using the \texttt{momentuHMM} package to be easily approachable and applicable. However, it is not currently possible to implement the circular-linear formulation described in Eq. \ref{eq:BCRW.TAm.alt}. The only way to estimate both strength of movement bias and direction of bias in \texttt{momentuHMM} is using the circular-circular link function. Implementing the model using a circular-linear formulation would require writing a custom HMM, which may not be accessible to most practitioners. For most use-cases, the circular-circular distribution is unlikely to lead to significant errors in state misclassification or bias mischaracterization. Generally, within a given state, the highest concentration of movement steps are close to $\vartheta_S$, while movement farther from $\vartheta_S$ is less common. Simultaneously, the non-linear artefacts of the circular-circular link function are most significant as at values of $\psi_t$ opposite $\vartheta_S$. As a result, the non-linear artefacts are unlikely to cause significant errors. We therefore believe the benefits of being able to use the straightforward and well-documented \texttt{momentuHMM} package exceed any errors that may arise at the edge-cases of the circular-circular distribution. Nevertheless, where possible, implementing the circular-linear formulation described in Eq. \ref{eq:BCRW.TAm.alt} would be preferable.

\section{Appendix: Wind Field Simulation} \label{appendix:B}
\setcounter{figure}{0}   
This Appendix describes the methods for simulating a wind field used in the two simulation studies.

First, a $150 \times 150$ Gaussian random field was generated (with autocorrelation range set to 150 and nugget effect set to 0.2), where each cell represented air pressure over a 1 km$^2$. The Pressure field was generated using the \texttt{NMLR} package (Sciaini et al. 2018); autocorrelation range and magnitude of variation were set to 100. To decrease the degree of noise, we applied a moving average over a $7 \times 7$ window using the \texttt{raster} package \citep{Hijmans2016PackageModeling}; this reduced the pressure field $P$ to a $94 \times 94$ grid. Second, the pressure gradient components, $u^{(pressure)}$ and $v^{(pressure)}$, were calculated following 
\begin{linenomath*}  \begin{align} 
\label{eq:u_v_gradient}
    u_{i,j}^{(pressure)} = P_{i,j-1} - P_{i,j+1} + \cos(\pi/4)(P_{i-1,j-1}  +  P_{i+1,j-1} -  P_{i-1,j+1} - P_{i+1,j+1})     \quad    i,j \in \{2,93\} \\
    v_{i,j}^{(pressure)} = P_{i+1,j} - P_{i-1,j} + \cos(\pi/4)(P_{i+1,j-1}  +  P_{i+1,j+1} -  P_{i-1,j-1} - P_{i-1,j+1})     \quad    i,j \in \{2,93\}
\end{align}  \end{linenomath*}
Third, $u_{i,j}^{(pressure)}$ and $v_{i,j}^{(pressure)}$ were re-scaled to the obtain to a maximum wind speed of 15 m s$^{-1}$ following
\begin{linenomath*} \begin{equation}  
    \label{eq:u_v_scale}
    \begin{bmatrix}
        u_{i,j}^{(wind)}\\
        v_{i,j}^{(wind)}
    \end{bmatrix}  
    = 15\cdot
    \begin{bmatrix}
        u_{i,j}^{(pressure)} / \max|u_{i,j}^{(pressure)}| \\
        v_{i,j}^{(pressure)} / \max|v_{i,j}^{(pressure)}| 
    \end{bmatrix}  
    \quad    i,j \in \{2,93\}
\end{equation}  \end{linenomath*}
Last, to obtain the final wind vectors, $u_{i,j}^{(wind)'}$ and $v_{i,j}^{(wind)'}$, a $-45\degree$ Coriolis rotation was applying using the following rotation matrix:
\begin{linenomath*} \begin{equation}  
    \label{eq:u_v_rotation}
    \begin{bmatrix}
        u_{i,j}^{(wind)'}\\
        v_{i,j}^{(wind)'}
    \end{bmatrix}  
    = 
    \begin{bmatrix}
        \cos(-\pi/4) & - \sin(-\pi/4)\\
        \sin(-\pi/4) & \cos(-\pi/4)
    \end{bmatrix}  
    \begin{bmatrix}
        u_{i,j}^{(wind)}\\
        v_{i,j}^{(wind)}
    \end{bmatrix}  
    \quad    i,j \in \{2,93\}
\end{equation}  \end{linenomath*}

\begin{figure}[H]
    \centering
    \includegraphics[scale = 0.75]{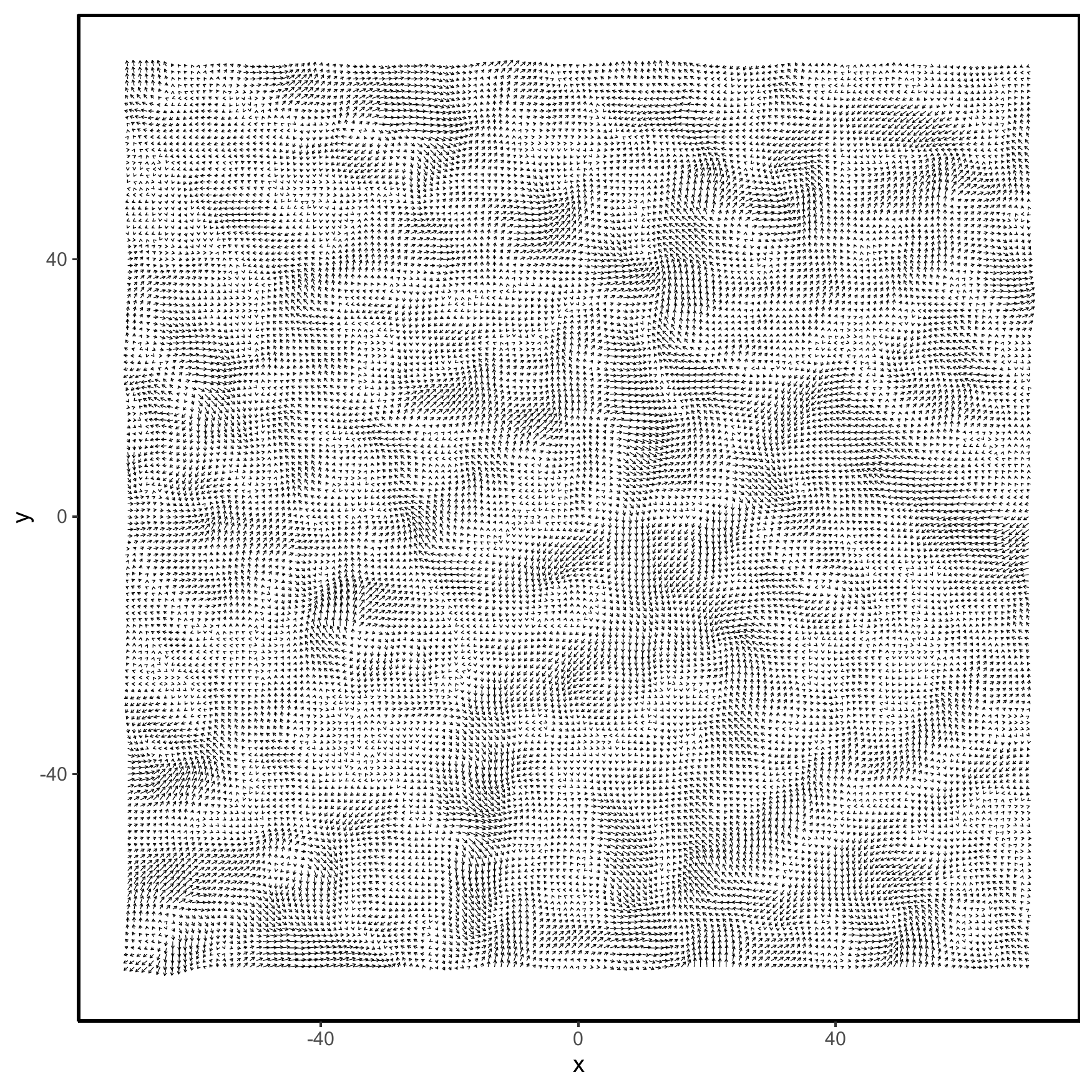}
    \caption{An example of a simulated wind field, where x and y are coordinates in metres. Vectors are derived from the simulated u and v components. Vector lengths are in m s$^{-1}$ and scaled by a factor of 0.10.}
    \label{fig:sim_wind}
\end{figure}

An example of a simulated wind field is presented in Fig. \ref{fig:sim_wind}.

\section{Appendix: simulation results} \label{appendix:C}
\setcounter{figure}{0}  
\setcounter{table}{0}  

\begin{table}[H]
    \centering
    \caption{Confusion matrix denoting the precision rate (bold diagonal; $true.state / (true.state + false.state$) and the false discovery rate (off-diagonal; $false.state / (true.state + false.state$) of the simulated 500 steps as predicted by the CRW HMM and BCRW HMM. Values denote the mean and [2.5\%, 97.5\%] quantiles across the 100 simulated tracks.}
    \begin{tabular}{@{}cccccc@{}} 
  &  &  & \multicolumn{3}{c}{Simulated state} \\ \cmidrule(l){4-6}  
  &  &  & D & O & ARS \\ \cmidrule(l){4-6}  
 \multicolumn{1}{l|}{\multirow{7}{*}{\rotatebox[origin=c]{90}{Predicted state}}} & \multicolumn{1}{l|}{\multirow{3}{*}{\begin{tabular}[c]{@{}c@{}}CRW\\HMM\end{tabular}}} & \multicolumn{1}{l|}{D}  & \textbf{ 0.37 [0.35, 0.39]} & 0.01 [0.00, 0.02] & 0.09 [0.06, 0.11]  \\ 
 \multicolumn{1}{l|}{} & \multicolumn{1}{l|}{} & \multicolumn{1}{l|}{O} & 0.06 [0.05, 0.07] & \textbf{ 0.95 [0.93, 0.96]} & 0.00 [0.00, 0.01]  \\ 
 \multicolumn{1}{l|}{} & \multicolumn{1}{l|}{} & \multicolumn{1}{l|}{ARS} & 0.57 [0.55, 0.59] & 0.04 [0.04, 0.05] & \textbf{ 0.91 [0.89, 0.93]}  \\ 
 \multicolumn{1}{l|}{} &  &  &  &  &  \\ 
 \multicolumn{1}{l|}{} & \multicolumn{1}{l|}{\multirow{3}{*}{\begin{tabular}[c]{@{}c@{}}BCRW\\HMM\end{tabular}}} & \multicolumn{1}{l|}{D} & \textbf{ 0.97 [0.97, 0.98]} & 0.00 [0.00, 0.00] & 0.06 [0.06, 0.07]  \\ 
 \multicolumn{1}{l|}{} & \multicolumn{1}{l|}{} & \multicolumn{1}{l|}{O} & 0.00 [0.00, 0.00] & \textbf{ 0.99 [0.98, 0.99]} & 0.01 [0.01, 0.01]   \\ 
 \multicolumn{1}{l|}{} & \multicolumn{1}{l|}{} & \multicolumn{1}{l|}{ARS} & 0.03 [0.02, 0.03] & 0.01 [0.01, 0.02] & \textbf{ 0.93 [0.93, 0.94]}  
 \end{tabular}
    \label{tab:Sim1_CM}
\end{table}

\label{appendix:D}


\section*{Supplementary material}

# D Appendix: Menotactic HMM Tutorial

## Introduction

In this R tutorial, we illustrate the implementation of the menotactic hidden Markov model described in the main text of Togunov *et al.* (2021) using the `momentuHMM` package (McClintock & Michelot, 2018). The tutorial is divided into three sections: **Data simulation**, **Model preparation and fit**, and **Model evaluation**.

We begin by first loading relevant packages and sourcing custom helper functions.

```
library(tidyverse)    # dplyr, ggplot2, & friends
library(NLMR)         # generate random environmental layers
library(raster)       # extent function for simulation
library(momentuHMM)   # run HMM
library(circular)     # fit von Mises curves
library(boot)         # logit and inverse logit functions
library(movMF)        # define circular von Mises distribution
library(cowplot)      # panel plots

source('Tutorial Source.R')
```

## Data simulation

As the foundation of this tutorial we will use a simulate track across a generated wind field. Such simulations allow us to assess model performance across multiple iterations and investigate the effects of different parameters (e.g., movement characteristics or degree of environmental error) or model structures (e.g., with different behaviours or behaviour associations).

### Simulate random wind field

We first simulate a vector field representing wind speed and directions as described in Togunov *et al.* (2021) Appendix A. We use the custom `sim_wind` function (available in the Source code of this Markdown) to simulate the wind field and extract the wind speed and direction:

```
wind_ls <- sim_wind(autocorr_range = 100, area = 100, max_spd = 15,
                    smooth_window = 7, cell_res = 1)
spatialCov <- list(wndspd = wind_ls[[2]],   # wind speed
                   wnddir = wind_ls[[3]])   # wind direction
```

`sim_wind` first simulates a $100 \times 100$ pressure field, then applies a $7 \times 7$ window moving average to smooth the field, then estimates wind speed and direction based on pressure differential between adjacent cells. This produces a $91 \times 91$ grid wind vector field in m s$^{-1}$ with a resolution representing 1 km. `sim_wind` returns a list of five elements corresponding to air pressure, wind speed, wind direction, u component of wind, and v component of wind. For the simulation, we only need wind speed and direction. Later, we will down-sample the u and v components to represent environmental error associated with a coarse resolution.

### Simulate animal track

Next, we simulate one animal movement track across the wind field we created. The track follows a 4-state hidden Markov model as described in section *Simulation study* of the main text (Togunov *et al.* 2021).

The four states are: drift ($D$; speed is determined by wind speed, and direction is 15º relative to wind), olfactory search left of wind ($O^{(L)}$; higher speed and 90º relative to wind), olfactory search right of wind ($O^{(R)}$; higher speed and $-90$º relative to wind), and area restricted search ($ARS$; slower speed and tortuous movement independent of wind). A detailed statistical descriptions of the model is provided in the main text (Togunov *et al.* 2021). We first define the distributions, `dist` associated with step length (gamma distribution) and turning angle (von Mises distribution), the state names, `stateNames` (optional), and the number of behavioural states, `nbStates`:

```
dist <- list(step = "gamma", angle = "vm")
stateNames <- c("D", "O(L)", "O(R)", "ARS")
nbStates <- length(stateNames)
```

**Step length parameterization**

Complex models can be developed in `momentuHMM` using pseudo-design matrices (DMs). In these matrices, the rows correspond to the parameters associated with the data stream distributions (e.g., mean and standard deviation, SD, for the gamma distribution used for step length). The number of rows in a DM must equal the number of states × the number of parameters associated with the distribution of the data stream. The order of the rows correspond to the order of the parameters as shown in table 2 of McClintock & Michelot (2018). In our case, step length follows a gamma distribution defined by a mean and SD parameters, and mean and concentration ($\kappa$) for turning angle angle. Thus, the first five rows of the of the step length DM correspond to the mean of the five states, and the following five rows represent the step length SD for the five states. All states are assumed to follow the same distribution, but the DM combined with fixing parameter coefficients allows us to develop complex and varried relationships among the states. In this study, the mean step length for drift, $\mu_{t,D}^{(l)}$, was a slope-intercept function of wind speed, $r_t$, following:

$$\ln(\mu_{t,D}^{(l)}) = \beta_{1,D} + \beta_{2,D} r_t \tag{1}$$

In the DM, the intercept, $\beta_{1,D}$, is represented by the '1' in the first column, and the slope corresponding to wind speed, $\beta_{2,D}$, is represented by `"wndspd"` in the second column. Both $O^{(L)}$ and $O^{(R)}$ share the parameters for mean and SD, so they share 1s in the same columns. The mean and SD for $D$ and $ARS$ are unique and therefore have their own columns. The design matrix for step length will look as follows:

```
# Model Design matrix & Parameters
#           D mean,  D slope~wind, O(L,R) mean, ARS mean, D SD, O(L,R) SD, ARS SD
stepDM <- matrix(c(1,       "wndspd",          0,          0,      0,     0, 0,  # m D
                   0,            0,            1,          0,      0,     0, 0,  # m O(L)
                   0,            0,            1,          0,      0,     0, 0,  # m O(R)
                   0,            0,            0,          1,      0,     0, 0,  # m ARS
                   0,            0,            0,          0,      1,     0, 0,  # sd D
                   0,            0,            0,          0,      0,     1, 0,  # sd O(L)
                   0,            0,            0,          0,      0,     1, 0,  # sd O(R)
                   0,            0,            0,          0,      0,     0, 1), # sd ARS
              nrow = nbStates*2, byrow = T,
              dimnames = list(c(paste0("m",1:nbStates),
                                paste0("sd",1:nbStates)),
                              c("m D intercept","m D slope", "m O(L,R)","m ARS",
                                "sd D", "sd O(L,R)", "sd ARS")))
```

Last, we define the coefficients corresponding to the columns of the step length DM to parameterise the step length for the simulation:

```r
stepPar <- c(-2.2, 0.1,          # m: intercept D, slope D
             0.1, -2.5,          # m: O(L,R), ARS
             -2.4, -0.5, -2.8)   # sd: D, O(L,R), ARS
```

**Turning angle parameterization**

We model mean turning angle $\mu_t^{(\phi)}$ at time $D^{(f)}$ with a bias relative stimulus $i$, using the circular-circular regression based on Rivest et al. (2016), which follows the general form:

$$\mu_t^{(\phi)} = \text{atan2}(\sum_{i=1}^{K} \alpha_i \sin \psi_{i,t}, 1 + \sum_{i=1}^{K} \alpha_i \cos \psi_{i,t}) \qquad (2)$$

where, $\psi_{i,t}$ represents the angle of the stimuli, $K$ represents the total number of stimuli, and $\alpha_i$ is the covariate-specific bias coefficient. To simulate a behaviour with bias toward any angle relative to a stimulus we model a movement with a bias parallel to the stimulus, $\alpha_1$, and a bias perpendicular to the stimulus, $\alpha_2$, following:

$$\mu_{S,t}^{(\phi)} = \frac{\text{atan2}(\alpha_{1,S} \sin(\psi_t) + \alpha_{2,S} \cos(\psi_t),}{1 + \alpha_{1,S} \cos(\psi_t) - \alpha_{2,S} \sin(\psi_t))} \qquad (3)$$

The `sin` and `cos` arguments of the circular-circular distributing in `momentuHMM` are fixed, so we must replace $\cos(\psi_t)$ in the numerator with $\sin(\psi_t + \pi/2)$ and replace $-\sin(\psi_t)$ in the denominator with $\cos(\psi_t + \pi/2)$. The model that will be fit in `momentuHMM` therefore follows:

$$\mu_{S,t}^{(\phi)} = \begin{cases} \text{atan2}(\alpha_{1,S} \sin(\psi_t) + \alpha_{2,S} \sin(\psi_t + \pi/2), & \text{if } S \in \{D, O^{(L)}\} \\ \quad 1 + \alpha_{1,S} \cos(\psi_t) + \alpha_{2,S} \cos(\psi_t + \pi/2)) & \\ \text{atan2}(\alpha_{1,S} \sin(\psi_t) + \alpha_{2,S} \sin(\psi_t - \pi/2), & \text{if } S = O^{(R)} \\ \quad 1 + \alpha_{1,S} \cos(\psi_t) + \alpha_{2,S} \cos(\psi_t - \pi/2)) & \\ \text{atan2}(0 \sin(\psi_t), 1 + 0 \cos(\psi_t)) & \text{if } S = ARS \end{cases} \qquad (4)$$

In `momentuHMM`, if any states have an angled covariate, this covariate must be included for each state in the DM. For states where there is no believed relationship (in this case, during ARS), the corresponding coefficients can be fixed to 0. The mean turning angle for $D$ is 15° relative to wind direction, `wnddir` and is simulated with a downwind bias (`"wnddir"` in the first column) and a cross-wind bias ("I(wnddir+pi/2)" in the second column). $O^{(L)}$ and $O^{(R)}$ are also simulated with downwind and cross-wind biases and share their coefficients (covariates in the same columns; note the '-' sign for $O^{(R)}$ representing bias right of wind).

```r
angleDM <- matrix(
#    D a1,      D a2              O(LR) a1,  O(LR) a2,          ARS a1,   D k,  O(LR) k,  ARS k
  c("wnddir","I(wnddir+pi/2)", 0,         0,                 0,        0,    0,        0,  # m D
    0,        0,               "wnddir","I(wnddir+pi/2)",0,           0,    0,        0,  # m O(L)
    0,        0,               "wnddir","I(wnddir-pi/2)",0,           0,    0,        0,  # m O(R)
    0,        0,               0,         0,                 "wnddir", 0,    0,        0,  # m ARS
    0,        0,               0,         0,                 0,        1,    0,        0,  # k D
    0,        0,               0,         0,                 0,        0,    1,        0,  # k O(L)
    0,        0,               0,         0,                 0,        0,    1,        0,  # k O(R)
    0,        0,               0,         0,                 0,        0,    0,        1), # k ARS
              nrow = nbStates*2, byrow = T,
              dimnames = list(c(paste0("m",1:nbStates),
                              paste0("k",1:nbStates)),
```

3

```
                            c("alpha_1 D","alpha_2 D",
                              "alpha_1 O(L,R)","alpha_2 O(L,R)",
                              "m ARS",
                              "k D", "k O(L,R)", "k ARS")))
```

Last, we define the coefficients corresponding to the columns of the turning angle DM:

```
anglePar <- c(100, -26.8,   # alpha_1, alpha_2: D
              0, 5,         # alpha_1, alpha_2: O(L,R)
              0,            # m: ARS
              5, 2, 0.5)    # k: D, O(L,R), ARS
```

**State transition probability parameterization**

The final parameters we must defined are the state transition probabilities. These are defined on the working scale (i.e., logit$\gamma_{i,j} \in (-\infty, \infty)$). The diagonal transition probabilities (i.e., probabilities of remaining in the same state) are used as a reference and are not explicitly defined. For legibility, we left blank space where the diagonal probabilities would lie.

```
# transition to:  D       O(L)    O(R)    ARS    # from:
beta <- matrix(c(         -2.7,   -2.7,   -3.9,  # D
                 -2.5,            -3.9,   -2.7,  # O(L)
                 -2.5,   -3.9,            -2.7,  # O(R)
                 -2.7,   -2.5,   -2.5           ),  # ARS
               nrow = 1)
```

The matrix coefficients of transition probabilities are defined in one row with additional rows representing covariates affecting transition probability (one row for each covariate; not applicable in our model) and columns representing off-diagonal elements of the transition probability matrix (TPM; McClintock & Michelot, 2018).

**Track simulation**

Finally, we simulate a single movement track of 500 steps using the `simData` function:

```
data <- simData(nbAnimals = 1,  obsPerAnimal = 500, nbStates = nbStates,
                stateNames = stateNames, dist = dist,
                DM = list(step = stepDM, angle = angleDM),
                Par = list(step = stepPar, angle = anglePar),
                beta = beta, spatialCovs = spatialCov, angleCovs = "wnddir",
                circularAngleMean = list(angle = T), states = TRUE, retrySims = 5)
```

`DM` represents the DMs used for the data streams, `Par` represents the values of the coefficients of the DMs, `circularAngleMean` indicates that the turning angle data stream follows the circular-circular distribution, and `angleCovs` indicates that the angle of wind direction is to be calculated relative to the animal bearing rather than the x-axis. `Par`, `DM`, and `circularAngleMean` (among other arguments not used in this model) require lists, with each named element corresponding to a data stream (i.e., "step" and "angle").

**Aggregate wind data**

Our model describes the use of a transition state to characterise biased random walks (BRWs) when faced with coarse environmental data. To illustrate this data loss, we use **aggregate** from the **raster** package to

decrease the resolution of the u and v components of wind (in this case from $1 \times 1$ km to $32 \times 32$ km) and then extract the wind data along the simulated track:

```
u <- aggregate(wind_ls[[4]], fact = 32) %>%
  raster::extract(as.matrix(data[,c("x","y")]), method = "bilinear")
v <- aggregate(wind_ls[[5]], fact = 32) %>%
  raster::extract(as.matrix(data[,c("x","y")]), method = "bilinear")
```

Last, we subset the recalculated wind speed and direction from the rarefied raster:

```
data.sub <- dplyr::select(data, "x", "y")
data.sub$wndspd <- sqrt(u^2 + v^2)
data.sub$wnddir <- atan2(v, u)
```

`data.sub` represents the data we may obtain from a GPS track after extracting imperfect wind data. The first six lines of the data are:

```
##              x         y   wndspd    wnddir
## 1  0.00000000 0.0000000 2.428804 0.6950117
## 2  0.03115136 0.6906565 2.262077 0.7163536
## 3 -0.23921255 1.1751329 2.168028 0.7249142
## 4 -0.10307737 1.2526750 2.139353 0.7315227
## 5  0.11465721 1.3510579 2.099712 0.7414846
## 6  0.18268340 1.4023982 2.082527 0.7454478
```

## Model preparation and fit

### Data preparation

To prepare the data for fitting a HMM, we use `prepData`, which calculates the step length and turning angle, from the movement track, along with the angle of wind relative to the animal bearing:

```
sim.data.prep <- prepData(data = data.sub, type = 'UTM', coordNames = c("x", "y"),
                          covNames = c("wnddir"), angleCovs = "wnddir")
```

### Model parameterization

We again define the distributions associated with each data stream, the state names, and number of states (in this case five, We divide $D$ into a 'first-step drift' state, $D^{(f)}$, and 'consecutive drift' state, $D^{(c)}$):

```
dist <- list(step = "gamma", angle = "vm")
stateNames <- c("D(f)", "D(c)", "O(L)", "O(R)", "ARS")
nbStates <- length(stateNames)
```

**Step length DM**

As in the simulation, we define the DMs for step length and turning angle. The matrix for step length is the same as in the simulation, with the addition of a row for mean and SD parameters for $D^{(f)}$. To share the step length parameters between $D^{(f)}$ and $D^{(c)}$, their columns share the same form (i.e., rows 1 and 2, and rows 6 and 7):

```r
stepDM <- matrix(c(1, "wndspd",0, 0, 0, 0, 0,   # m D(f)
                   1, "wndspd",0, 0, 0, 0, 0,   # m D(c)
                   0, 0,       1, 0, 0, 0, 0,   # m O(L)
                   0, 0,       1, 0, 0, 0, 0,   # m O(R)
                   0, 0,       0, 1, 0, 0, 0,   # m ARS
                   0, 0,       0, 0, 1, 0, 0,   # sd D(f)
                   0, 0,       0, 0, 1, 0, 0,   # sd D(c)
                   0, 0,       0, 0, 0, 1, 0,   # sd O(L)
                   0, 0,       0, 0, 0, 1, 0,   # sd O(R)
                   0, 0,       0, 0, 0, 0, 1),  # sd ARS
                 nrow = nbStates*2, byrow = T,
                 dimnames = list(c(paste0("m",1:nbStates),
                                   paste0("sd",1:nbStates)),
                                 c("m intercept D(f,c)", "m slope D(f,c)","m O(L,R)","m ARS",
                                   "sd D(f,c)", "sd O(L,R)", "sd ARS")))
```

`momentuHMM` allows us to add fix parameters using the `fixPar` argument, which takes a named list, where each element represents a vector of coefficient values corresponding to the data streams and the off-diagonal probabilities of the state TPM. When using a DM, the elements of the vector correspond to the columns of the matrix. Any values that are to be estimated are represented by an `NA` in `fixPar` argument. For step length, we want all coefficients to be estimated, thus we define this vector as seven `NA`:

```r
stepF <- rep(NA, 7)
```

**Turning angle DM**

We extend eq. 4 to include an additional function for the mean turning angle for $D^{(f)}$ following:

$$\mu_{S,t}^{(\phi)} = \begin{cases} \text{atan2}(100\sin(\psi_t) + \alpha_{2,S}\sin(\psi_t + \pi/2), & \text{if } S = D^{(f)} \\ \quad 1 + 100\cos(\psi_t) + \alpha_{S,2}\cos(\psi_t + \pi/2)) & \\ \text{atan2}(\alpha_{1,S}\sin(\psi_t) + \alpha_{2,S}\sin(\psi_t + \pi/2), & \text{if } S \in \{D^{(c)}, O^{(L)}\} \\ \quad 1 + \alpha_{1,S}\cos(\psi_t) + \alpha_{2,S}\cos(\psi_t + \pi/2)) & \\ \text{atan2}(\alpha_{1,S}\sin(\psi_t) + \alpha_{2,S}\sin(\psi_t - \pi/2), & \text{if } S = O^{(R)} \\ \quad 1 + \alpha_{1,S}\cos(\psi_t) + \alpha_{2,S}\cos(\psi_t - \pi/2)) & \\ \text{atan2}(0\sin(\psi_t), 1 + 0\cos(\psi_t)) & \text{if } S = ARS \end{cases} \quad (5)$$

The DM for this model is represented by:

```r
angleDM <- matrix(c("wnddir","I(wnddir + pi/2)",0,0,0,0,0,0,0,0,0,   # m D(f)
                    0,0,"wnddir","I(wnddir + pi/2)",0,0,0,0,0,0,0,   # m D(c)
                    0,0,0,0,"wnddir","I(wnddir + pi/2)",0,0,0,0,0,   # m O(L)
                    0,0,0,0,"wnddir","I(wnddir - pi/2)",0,0,0,0,0,   # m O(R)
                    0,0,0,0,0,0,"wnddir",0,0,0,0,    # m ARS
                    0,0,0,0,0,0,0,1,0,0,0,    # k D(f)
                    0,0,0,0,0,0,0,0,1,0,0,    # k D(c)
                    0,0,0,0,0,0,0,0,0,1,0,    # k O(L)
                    0,0,0,0,0,0,0,0,0,1,0,    # k O(R)
                    0,0,0,0,0,0,0,0,0,0,1),   # k ARS
                  nrow = nbStates*2, byrow = T,
                  dimnames = list(c(paste0("m",1:nbStates),
```

6

```
                                paste0("k",1:nbStates)),
                         c("alpha_1 D(f)","alpha_2 D(f)",
                           "alpha_1 D(c)","alpha_2 D(c)",
                           "alpha_1 O(L,R)","alpha_2 O(L,R)",
                           "alpha_1 ARS",
                           "k D(f)", "k D(c)", "k O(L,R)", "k ARS")))
```

For turning angle, $D^{(f)}$ and $D^{(c)}$ do not share any parameters to allow the model to determine the strength of bias relative to wind, given the error in the wind data. Although $D^{(f)}$ and $D^{(c)}$ represent the same behaviour, the low resolution of wind data may lead the model to estimate $D^{(c)}$ as an unbiased correlated random walk (CRW; see main text for details).

To constrain $D^{(f)}$ to be a BRW relative to wind, we fix $\alpha_{1,D^{(f)}}$ to a large number (in this case, 100). We must also fix the mean turning angle parameters for $ARS$ to 0 to constrain it to a CRW. All other parameters are to be estimated. $\alpha_{1,D^{(f)}}$ is represented by the first column of pseudo design matrix, and $\alpha_{1,ARS}$ is represented by the seventh column, the vector of fixed coefficients for turning angle follows:

```
angleF <- c(100, rep(NA, 5), 0, rep(NA,4))
```

**Transition probability matrix**

In addition to sharing data stream parameters, we shared the transition probability beta coefficients in the TPM to and from $O^{(L)}$ and $O(R)$. Such constraints on the beta coefficients are achieved using the `betaCons` argument. `betaCons` takes a matrix, where the first row corresponds to the off-diagonal elements of the TPM, and subsequent rows correspond to parameters associated with covariates (not applicable in our case). Shared parameters are indicated by matching numbers representing which index is being shared. Sharing transitions coefficients to and from $O^{(L)}$ and $O^{(R)}$ is represented by:

```
# transition to:    D(f) D(c) O(L) O(R) ARS    # from:
betaCons <- matrix(c(   1,   2,   2,   4,  # D(f)
                        5,        6,   6,   8,  # D(c)
                        9,  10,       11,  12,  # O(L)
                        9,  10,  11,       12,  # O(R)
                       17,  18,  19,  19      ), # ARS
            nrow = 1, ncol = 20, byrow = TRUE,
            dimnames = list(c("Intercept"),
                       c(              "D(f)->D(c)","D(f)->O(L)","D(f)->O(R)","D(f)->ARS",
                         "D(c)->D(f)",              "D(c)->O(L)","D(c)->O(R)","D(c)->ARS",
                         "O(L)->D(f)","O(L)->D(c)",             "O(L)->O(R)","O(L)->ARS",
                         "O(R)->D(f)","O(R)->D(c)","O(R)->O(L)",             "O(R)->ARS",
                         "ARS ->D(f)","ARS ->D(c)", "ARS->O(L)", "ARS->O(R)"          )))
```

Note: the index of any element (i.e., transition probability) that is using (i.e., sharing) the value of another index must be excluded from the rest of the matrix. In our case, the third element shares the parameter estimated for the second element, therefore both have the number 2 and the index of 3 cannot be referenced in the rest of the matrix. Diagonal values of the TPM are not estimated, however, we leave blank space where they would lie for legibility.

To ensure $D^{(f)}$ to be a one-step transition to $D^{(c)}$, we impose constraints on the TPM using the `fixPar` argument. For fixed transition probabilities, `fixPar` takes a vector corresponding to the off-diagonal elements of the TPM. Values must be provided on the working scale, where probabilities of 0 are represented by very large negative values and probabilities of 1 are represented by very large positive values. Here, the transition from $D^{(f)}$ to $D^{(c)}$ was set to 200 (first element), and transitions from $D^{(f)}$ to all other states was set to

-200 (elements 2-4). The transition from $D^{(c)}$ to $D^{(f)}$ was set to -200 (5th element), and transitions to $D^{(c)}$ from $O^{(L)}$, $O^{(R)}$, and $ARS$ were set to -200 (elements 10, 14, and 18, respectively). All other transition probabilities are to be estimated and are represented by `NA`:

```r
# trans' to: D(f)  D(c)   O(L)  O(R)   ARS   # from:
betaF <- c(        200, -200, -200, -200, # D(f)
            -200,         NA ,  NA ,  NA, # D(c)
             NA , -200,         NA ,  NA, # O(L)
             NA , -200,  NA ,         NA, # O(R)
             NA , -200,  NA ,  NA       ) # ARS
```

Note: we leave blank space where the diagonal probabilities would lie for legibility.

**Starting parameters**

The last argument necessary to run the model is the starting parameters from which the `momentuHMM` will begin optimization. The minimum required starting parameters are vectors correspond to the columns of the DMs. These can be obtained through data exploration and developing simpler HMMs with fewer states.

```r
step0 <- c(-2.2, 0.1, 0.1, -2.5,  # m: intercept D(f) & D(c), slope D(f) & D(c), O(L,R), ARS
           -2.4, -0.5, -2.8)      # SD: D(f) & D(c), O(L,R), ARS

angle0 <- c(100, -26.8,   # alpha_1, alpha_2: D(f)
            100, -26.8,   # alpha_1, alpha_2: D(c)
            0, 5,         # alpha_1, alpha_2: O(L,R)
            0,            # alpha_1: ARS
            5, 5, 2, 0.5) # k: D(f), D(c), O(L,R), ARS
```

**Fit 5-state HMM**

Finally, we can fit the model described in the main text using the `fitHMM` function:

```r
fit_HMM <- fitHMM(data = sim.data.prep, nbStates = nbStates, stateNames = stateNames,
                  dist = dist, DM = list(angle = angleDM, step = stepDM),
                  fixPar = list(step = stepF, angle = angleF, beta = betaF),
                  betaCons = betaCons, Par0 = list(step = step0, angle = angle0),
                  estAngleMean = list(angle = TRUE), circularAngleMean = list(angle = TRUE))
```

`fixPar` represents the fixed parameters/coefficients for the data streams and TMP, `betaCons` represents the constraints on the TPM on which parameters are shared, `Par0` represents the starting parameters from which optimization is initiated. Other arguments have been previousy defined.

# Model evaluation

We can decode the most likely state sequence using the 'viterbi' function:

```r
sim.data.prep$states <- viterbi(fit_HMM)
```

Fig. 1 plots a simulated track coloured by the decoded states overlaid on the wind field.

```r
# define state name & colours
state_names <- c(bquote(D^(f)), bquote(D^(c)), bquote(O^(L)), bquote(O^(R)), "ARS")
state_cols <- data.frame(states = 1:nbStates,
        colour = c("#3770F2", "#CADAFD", "#FFAE00", "#EB4334", "#1F8740"))
state_cols$colour <- factor(state_cols$colour, levels = state_cols$colour)

# join decoded states with colour
sim.data.prep <- left_join(sim.data.prep, state_cols, by = "states")

# plot decoded track
ggplot(sim.data.prep, aes(x = x, y = y, colour = colour, group = 1), parse = T) + geom_path() +
  scale_color_identity(labels = state_names, guide = "legend") +
  xlim(c(min(data$x) - 1, max(data$x) + 1)) + ylim(c(min(data$y) - 1, max(data$y) + 1)) +
  labs(colour = "State") + coord_fixed() + theme_classic()
```

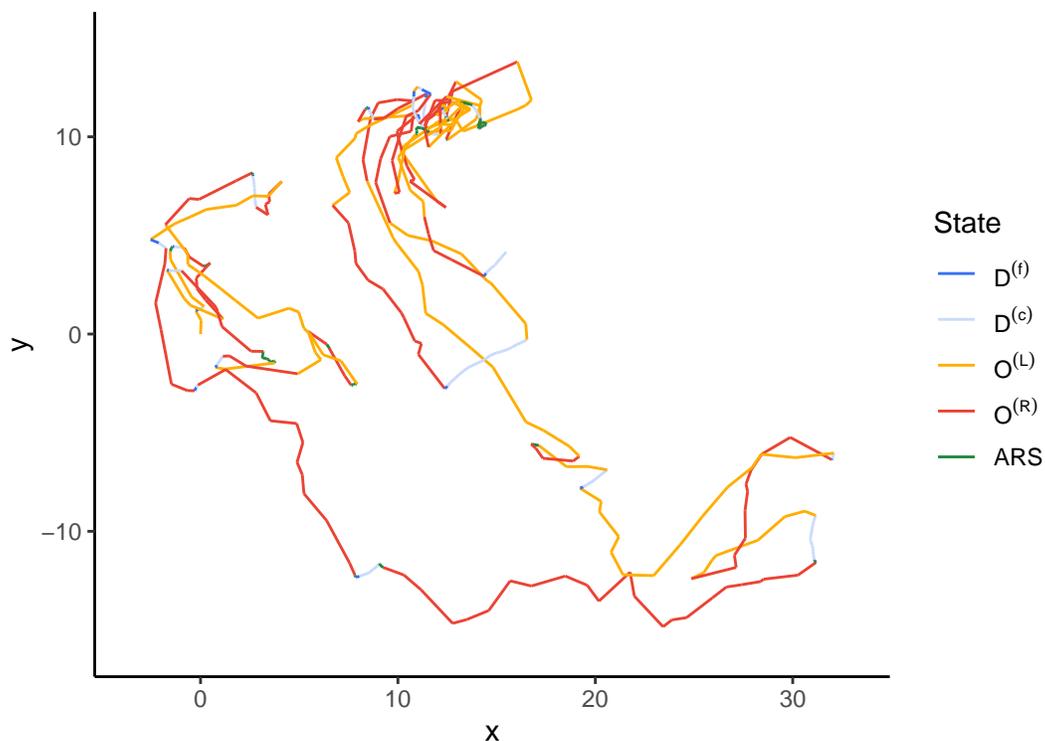

Figure 1: Simulated polar bear track overlayed on the simulated wind field. The track colours represent the predicted states of a five-state HMM determined using the Viterbi algorithm.

## Parameter estimation

The estimated state TPM, Γ, can be called directly from the fit model - the values are on the working scale and can be converted to true probabilities using `boot::inv.logit`:

```
fit_HMM$mle$gamma %>%
  inv.logit() %>% round(2)
```

```
##     D(f) D(c) O(L) O(R)  ARS
## D(f) 0.50 0.73 0.50 0.50 0.50
## D(c) 0.50 0.69 0.52 0.52 0.51
## O(L) 0.51 0.50 0.69 0.53 0.51
## O(R) 0.51 0.50 0.53 0.69 0.51
## ARS  0.53 0.50 0.51 0.51 0.68
```

Given the TPM, we can calculate the estimated time spent in each state as predicted by the stationary state distribution.

Using the `stationary`.

```
stationary <- stationary(fit_HMM)
stationary
```

```
## [[1]]
##            D(f)       D(c)       O(L)       O(R)       ARS
## [1,] 0.04967745 0.2443671 0.2642635 0.2642635 0.1774284
```

Our model predicts the animal spent 5% in $D^{(f)}$, 24.4% in $D^{(c)}$, 26.4% in $O^{(L)}$, 26.4% in $O^{(R)}$, and 17.7% in $ARS$.

Of particular importance in our model is the state-specific angle of attraction relative to stimulus that is given by:
$$\vartheta_S = \operatorname{atan2}(\alpha_{2,S}, \alpha_{1_S}) \tag{6}$$

Our model estimated $\vartheta_{D^{(f)}} = -10.7°$, $\vartheta_{D^{(c)}} = -154.3°$, $\vartheta_{O^{(L)}} = 91.3°$, and $\vartheta_{O^{(R)}} = -91.3°$.

To classify the behaviours as biased (BRW), correlated (CRW), or biased-correlated random walks (BCRW), we calculate the scaled magnitude of attraction, $M_S^*$, represented by:

$$M_S^* = \frac{\sqrt{\alpha_{1,S}^2 + \alpha_{2,S}^2}}{1 + \sqrt{\alpha_{1,S}^2 + \alpha_{2,S}^2}} \tag{7}$$

```
# calculate magnitude of attraction
Mag_att <- function(alpha_1, alpha_2){
  sqrt(alpha_1^2 + alpha_2^2)/(1 + sqrt(alpha_1^2 + alpha_2^2))
}
```

In this case, the magnitude of attraction for $D^{(f)}$ at was $M_{D^{(f)}}^* = 0.99$, consistent with a BRW; for $D^{(c)}$ was $M_{D^{(c)}}^* = 0.01$, consistent with a CRW; and for $O^{(L,R)}$ was $M_{O^{(L,R)}}^* = 0.55$, consistent with a BCRW.

### Data stream distributions

We can visualise the estimated step length distribution (Fig. 2) and turning angle distribution (Fig. 3) among the different states.

When covariates are present in the DM, we must specify the value at which to predict the distribution. We assume wind speed is equal to the median observed along the track (i.e., $r_t = 2$ m s$^{-1}$), and the relative direction of wind is in the same direction as movement (i.e., $\psi_t = 0°$).

```r
######### Step length ########## #
# define design-matrix columns associated with each coefficient
Beta_coln <- data.frame(state = 1:5,
                        Beta_mean_i = c(1, 1, 3, 3, 4),    # step length intercept
                        Beta_mean_s = c(2, 2, NA, NA, NA), # slope ~ wind speed
                        Beta_sd = c(5, 5, 6, 6, 7))        # step length SD
cov <- round(median(fit_HMM$data$wndspd))   # median value for wind speed as covariate

# get the estimated step length parameters from the fit HMM
Par <- getPar(fit_HMM)$Par$step
for(i in 2:ncol(Beta_coln)){
    Beta_coln[,i] <- Par[Beta_coln[,i]]
    Beta_coln[is.na(Beta_coln[,i]),i] <- 0
}
# get max step length as upper limit for distribution
max_step <- ceiling(max(fit_HMM$data$step, na.rm = T))
x <- seq(0.001, max_step*1.1, max_step/1000)
# predict Density based on model parameters
df <- NULL
for(i in 1:nrow(Beta_coln)){
    mean <- exp(Beta_coln[i,2] + Beta_coln[i,3]*cov)
    sd <- exp(Beta_coln[i,4])
    y <- dgamma(x = x, shape = (mean/sd)^2, rate = mean/(sd)^2)*
      stationary(fit_HMM)[[1]][Beta_coln[i,1]]  # scale on stationary state distribution
  df <- rbind(df, data.frame(states = i, x, y, data_stream = "Step length"))
}

######### Turning angle ########## #
Beta_coln <- data.frame(state = 1:5,
                        alpha_1 = c(1, 3, 5, 5, 7),    # bias downwind
                        alpha_2 = c(2, 4, 6, -6, NA),  # bias cross-wind
                        kappa = c(8, 9, 10, 10, 11))   # Turning angle concentration
cov <- 0
mirror <- ifelse(Beta_coln < 0, -1, 1)  # take negative of estimate for shared parameters
Beta_coln <- abs(Beta_coln)
# get the estimated turning angle parameters from the fit HMM
Beta_coln[,2] <- fit_HMM$mle$angle[Beta_coln[,2]]*mirror[,2]
Beta_coln[,3] <- fit_HMM$mle$angle[Beta_coln[,3]]*mirror[,3]
Beta_coln[is.na(Beta_coln[,3]),3] <- 0
Beta_coln[,4] <- exp(fit_HMM$mle$angle[Beta_coln[,4]])*mirror[,4]
x <- seq(-pi,pi, 0.01)
# predict Density
for(i in 1:nrow(Beta_coln)){
     alpha_1 = Beta_coln[i,2]; alpha_2 = Beta_coln[i,3]
     y <- dvonmises(x = x, mu = atan2(alpha_1*sin(cov) + alpha_2*cos(cov),
                                     1 + alpha_1*cos(cov) - alpha_2*sin(cov)),
                    kappa = Beta_coln[i,4])*stationary(fit_HMM)[[1]][Beta_coln[i,1]]
     df <- rbind(df, data.frame(states = i, x, y, data_stream = "Turning angle"))
  }
df <- df %>% left_join(state_cols, by = "states")  # plot state curves

# plot state density curves # line plot coloured by state
density_plot <- ggplot(df, aes(x = x, y = y, col = colour)) + geom_line() +
```

```
  scale_color_identity(labels = state_names, guide = "legend") +
  xlab(NULL) + ylab("Density") + labs(colour = "States") +  # labels
  facet_wrap(~data_stream, scales = "free", strip.position = "bottom") +  # facet by DS
  theme(strip.background = element_blank(), strip.placement = "outside")
density_plot
```

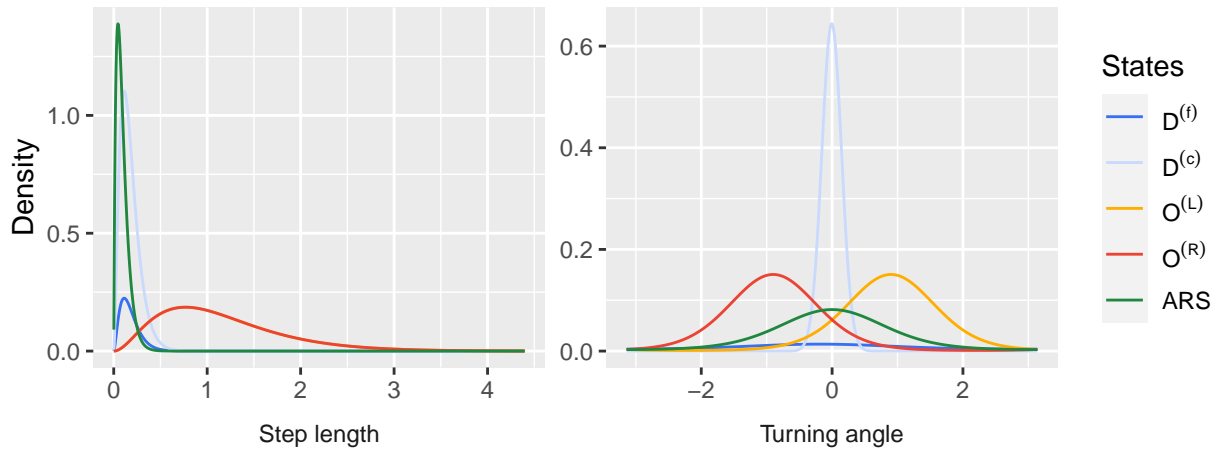

Figure 2: Probability density plots based on a fitted five-state HMM for step length $l_t$ (km 30min$^{-1}$) assuming wind speed $r_t = 2\,\text{m}\,\text{s}^{-1}$ and turning angle $\phi_t$ (in radians) assuming relative angle $\psi_t = 0$. The height of each density distribution is scaled by the stationary state distribution of each respective state.

### Turning angle prediction

Last, we may be interested in the predicted turning angle, $\phi_t$ as a function of underlying relative wind direction, $\psi_t$ (Fig. 3).

```r
# plot psi for D(f) and D(c)
p_psi_D <- plot_psi(model = fit_HMM, data = sim.data.prep, state_names = stateNames,
                    state_no = c(1,2), colours = state_cols[,2],
                    beta_coln = c(1,2,3,4))
# plot psi for O(L) and O(R)
alpha_1 <- fit_HMM$mle$angle[5]
alpha_2 <- fit_HMM$mle$angle[6]
data_2 <- data.frame(x = seq(-pi,pi,0.01),
                     y = atan2(alpha_1*sin(seq(-pi,pi,0.01)) -
                                 alpha_2*sin(seq(-pi,pi,0.01)+pi/2),
                               1 + alpha_1*cos(seq(-pi,pi,0.01)) -
                                 alpha_2*cos(seq(-pi,pi,0.01)+pi/2)))

p_psi_O <- plot_psi(model = fit_HMM, data = sim.data.prep, state_names = stateNames,
                    state_no = c(3,4), colours = state_cols[,2],
                    beta_coln = c(5,6)) +
  geom_line(data = data_2, aes(x = x, y = y),
            col = state_cols[4,2], size = 0.7, alpha = 0.8)

p_psi_A <- plot_psi(model = fit_HMM, data = sim.data.prep, state_names = stateNames,
```

```
                        state_no = c(5), colours = state_cols)
# combine psi plots and add legend
plot_grid(p_psi_D, p_psi_O, p_psi_A, get_legend(density_plot), nrow = 1,
          labels = c("(a)", "(b)", "(c)", ""), rel_widths = c(1,1,1, .3))
```

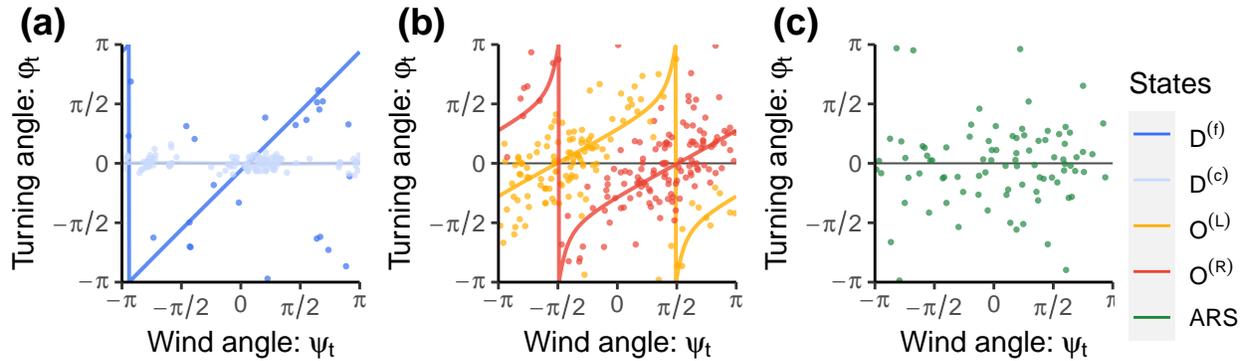

Figure 3: Turning angle $\phi_t$ (in radians) versus angle of wind relative to previous step $\psi_t$ (in radians). Lines represent the expected turning angle based on the estimated values of $\alpha_1$ and $\alpha_2$, given $\psi_t$.



\end{document}